\documentclass[twocolumn,letterpaper,aps,prc,superscriptaddress,showpacs,nofootinbib,floatfix]{revtex4}

\usepackage{graphicx}	
\usepackage{xspace}     

\newcommand{\Npart}{\mbox{$N_{\rm part}$}\xspace}
\newcommand{\Ncoll}{\mbox{$N_{\rm coll}$}\xspace}
\newcommand{\Nch}{\mbox{$N_{\rm ch}$}\xspace}
\newcommand{\Et}{\mbox{${\rm E}_T$}\xspace}

\newcommand{\sqs}{\mbox{$\sqrt{s}$}\xspace}
\newcommand{\sqsn}{\mbox{$\sqrt{s_{_{NN}}}$}\xspace}
\newcommand{\Etemc}{\mbox{${\rm{E}}_{T\,{\rm EMC}}$}\xspace}
\newcommand{\Nqp}{\mbox{$N_{qp}$}\xspace}

\def\lsim{\raise0.3ex\hbox{$<$\kern-0.75em\raise-1.1ex\hbox{$\sim$}}}
\def\gsim{\raise0.3ex\hbox{$>$\kern-0.75em\raise-1.1ex\hbox{$\sim$}}}
\def\mean#1{\langle#1\rangle}

\begin{document}

\title{Transverse-energy distributions at midrapidity in $p$$+$$p$, 
$d$$+$Au, and Au$+$Au collisions at $\sqrt{s_{_{NN}}}=62.4$--200~GeV and 
implications for particle-production models.}

\newcommand{\abilene}{Abilene Christian University, Abilene, Texas 79699, USA}
\newcommand{\acadsin}{Institute of Physics, Academia Sinica, Taipei 11529, Taiwan}
\newcommand{\augie}{Department of Physics, Augustana College, Sioux Falls, South Dakota 57197, USA}
\newcommand{\banaras}{Department of Physics, Banaras Hindu University, Varanasi 221005, India}
\newcommand{\barc}{Bhabha Atomic Research Centre, Bombay 400 085, India}
\newcommand{\baruch}{Baruch College, City University of New York, New York, New York, 10010 USA}
\newcommand{\bnlphys}{Brookhaven National Laboratory, Upton, New York 11973-5000, USA}
\newcommand{\caucr}{University of California - Riverside, Riverside, California 92521, USA}
\newcommand{\ciae}{Science and Technology on Nuclear Data Laboratory, China Institute of Atomic Energy, Beijing 102413, P.~R.~China}
\newcommand{\cns}{Center for Nuclear Study, Graduate School of Science, University of Tokyo, 7-3-1 Hongo, Bunkyo, Tokyo 113-0033, Japan}
\newcommand{\colorado}{University of Colorado, Boulder, Colorado 80309, USA}
\newcommand{\columbia}{Columbia University, New York, New York 10027 and Nevis Laboratories, Irvington, New York 10533, USA}
\newcommand{\dapnia}{Dapnia, CEA Saclay, F-91191, Gif-sur-Yvette, France}
\newcommand{\debrecen}{Debrecen University, H-4010 Debrecen, Egyetem t{\'e}r 1, Hungary}
\newcommand{\elte}{ELTE, E{\"o}tv{\"o}s Lor{\'a}nd University, H - 1117 Budapest, P{\'a}zm{\'a}ny P. s. 1/A, Hungary}
\newcommand{\fsu}{Florida State University, Tallahassee, Florida 32306, USA}
\newcommand{\gsu}{Georgia State University, Atlanta, Georgia 30303, USA}
\newcommand{\hiroshima}{Hiroshima University, Kagamiyama, Higashi-Hiroshima 739-8526, Japan}
\newcommand{\ihepprot}{IHEP Protvino, State Research Center of Russian Federation, Institute for High Energy Physics, Protvino, 142281, Russia}
\newcommand{\illuiuc}{University of Illinois at Urbana-Champaign, Urbana, Illinois 61801, USA}
\newcommand{\inrras}{Institute for Nuclear Research of the Russian Academy of Sciences, prospekt 60-letiya Oktyabrya 7a, Moscow 117312, Russia}
\newcommand{\isu}{Iowa State University, Ames, Iowa 50011, USA}
\newcommand{\jaea}{Advanced Science Research Center, Japan Atomic Energy Agency, 2-4 Shirakata Shirane, Tokai-mura, Naka-gun, Ibaraki-ken 319-1195, Japan}
\newcommand{\jinrdubna}{Joint Institute for Nuclear Research, 141980 Dubna, Moscow Region, Russia}
\newcommand{\kaeri}{KAERI, Cyclotron Application Laboratory, Seoul, Korea}
\newcommand{\kek}{KEK, High Energy Accelerator Research Organization, Tsukuba, Ibaraki 305-0801, Japan}
\newcommand{\korea}{Korea University, Seoul, 136-701, Korea}
\newcommand{\kurchatov}{Russian Research Center ``Kurchatov Institute", Moscow, 123098 Russia}
\newcommand{\kyoto}{Kyoto University, Kyoto 606-8502, Japan}
\newcommand{\labllr}{Laboratoire Leprince-Ringuet, Ecole Polytechnique, CNRS-IN2P3, Route de Saclay, F-91128, Palaiseau, France}
\newcommand{\lahorelums}{Physics Department, Lahore University of Management Sciences, Lahore, Pakistan}
\newcommand{\lawllnl}{Lawrence Livermore National Laboratory, Livermore, California 94550, USA}
\newcommand{\losalamos}{Los Alamos National Laboratory, Los Alamos, New Mexico 87545, USA}
\newcommand{\lpc}{LPC, Universit{\'e} Blaise Pascal, CNRS-IN2P3, Clermont-Fd, 63177 Aubiere Cedex, France}
\newcommand{\lund}{Department of Physics, Lund University, Box 118, SE-221 00 Lund, Sweden}
\newcommand{\michigan}{Department of Physics, University of Michigan, Ann Arbor, Michigan 48109-1040, USA}
\newcommand{\muenster}{Institut f\"ur Kernphysik, University of Muenster, D-48149 Muenster, Germany}
\newcommand{\myongji}{Myongji University, Yongin, Kyonggido 449-728, Korea}
\newcommand{\nagasaki}{Nagasaki Institute of Applied Science, Nagasaki-shi, Nagasaki 851-0193, Japan}
\newcommand{\newmex}{University of New Mexico, Albuquerque, New Mexico 87131, USA }
\newcommand{\nmsu}{New Mexico State University, Las Cruces, New Mexico 88003, USA}
\newcommand{\ohio}{Department of Physics and Astronomy, Ohio University, Athens, Ohio 45701, USA}
\newcommand{\ornl}{Oak Ridge National Laboratory, Oak Ridge, Tennessee 37831, USA}
\newcommand{\orsay}{IPN-Orsay, Universite Paris Sud, CNRS-IN2P3, BP1, F-91406, Orsay, France}
\newcommand{\peking}{Peking University, Beijing 100871, P.~R.~China}
\newcommand{\pnpi}{PNPI, Petersburg Nuclear Physics Institute, Gatchina, Leningrad region, 188300, Russia}
\newcommand{\riken}{RIKEN Nishina Center for Accelerator-Based Science, Wako, Saitama 351-0198, Japan}
\newcommand{\rikjrbrc}{RIKEN BNL Research Center, Brookhaven National Laboratory, Upton, New York 11973-5000, USA}
\newcommand{\rikkyo}{Physics Department, Rikkyo University, 3-34-1 Nishi-Ikebukuro, Toshima, Tokyo 171-8501, Japan}
\newcommand{\saispbstu}{Saint Petersburg State Polytechnic University, St. Petersburg, 195251 Russia}
\newcommand{\saopaulo}{Universidade de S{\~a}o Paulo, Instituto de F\'{\i}sica, Caixa Postal 66318, S{\~a}o Paulo CEP05315-970, Brazil}
\newcommand{\seoulnat}{Seoul National University, Seoul, Korea}
\newcommand{\stonybrkc}{Chemistry Department, Stony Brook University, SUNY, Stony Brook, New York 11794-3400, USA}
\newcommand{\stonycrkp}{Department of Physics and Astronomy, Stony Brook University, SUNY, Stony Brook, New York 11794-3400, USA}
\newcommand{\subatech}{SUBATECH (Ecole des Mines de Nantes, CNRS-IN2P3, Universit{\'e} de Nantes) BP 20722 - 44307, Nantes, France}
\newcommand{\tenn}{University of Tennessee, Knoxville, Tennessee 37996, USA}
\newcommand{\titech}{Department of Physics, Tokyo Institute of Technology, Oh-okayama, Meguro, Tokyo 152-8551, Japan}
\newcommand{\tsukuba}{Institute of Physics, University of Tsukuba, Tsukuba, Ibaraki 305, Japan}
\newcommand{\vandy}{Vanderbilt University, Nashville, Tennessee 37235, USA}
\newcommand{\waseda}{Waseda University, Advanced Research Institute for Science and Engineering, 17 Kikui-cho, Shinjuku-ku, Tokyo 162-0044, Japan}
\newcommand{\weizmann}{Weizmann Institute, Rehovot 76100, Israel}
\newcommand{\wigner}{Institute for Particle and Nuclear Physics, Wigner Research Centre for Physics, Hungarian Academy of Sciences (Wigner RCP, RMKI) H-1525 Budapest 114, POBox 49, Budapest, Hungary}
\newcommand{\yonsei}{Yonsei University, IPAP, Seoul 120-749, Korea}
\affiliation{\abilene}
\affiliation{\acadsin}
\affiliation{\augie}
\affiliation{\banaras}
\affiliation{\barc}
\affiliation{\baruch}
\affiliation{\bnlphys}
\affiliation{\caucr}
\affiliation{\ciae}
\affiliation{\cns}
\affiliation{\colorado}
\affiliation{\columbia}
\affiliation{\dapnia}
\affiliation{\debrecen}
\affiliation{\elte}
\affiliation{\fsu}
\affiliation{\gsu}
\affiliation{\hiroshima}
\affiliation{\ihepprot}
\affiliation{\illuiuc}
\affiliation{\inrras}
\affiliation{\isu}
\affiliation{\jaea}
\affiliation{\jinrdubna}
\affiliation{\kaeri}
\affiliation{\kek}
\affiliation{\korea}
\affiliation{\kurchatov}
\affiliation{\kyoto}
\affiliation{\labllr}
\affiliation{\lahorelums}
\affiliation{\lawllnl}
\affiliation{\losalamos}
\affiliation{\lpc}
\affiliation{\lund}
\affiliation{\michigan}
\affiliation{\muenster}
\affiliation{\myongji}
\affiliation{\nagasaki}
\affiliation{\newmex}
\affiliation{\nmsu}
\affiliation{\ohio}
\affiliation{\ornl}
\affiliation{\orsay}
\affiliation{\peking}
\affiliation{\pnpi}
\affiliation{\riken}
\affiliation{\rikjrbrc}
\affiliation{\rikkyo}
\affiliation{\saispbstu}
\affiliation{\saopaulo}
\affiliation{\seoulnat}
\affiliation{\stonybrkc}
\affiliation{\stonycrkp}
\affiliation{\subatech}
\affiliation{\tenn}
\affiliation{\titech}
\affiliation{\tsukuba}
\affiliation{\vandy}
\affiliation{\waseda}
\affiliation{\weizmann}
\affiliation{\wigner}
\affiliation{\yonsei}
\author{S.S.~Adler} \affiliation{\bnlphys}
\author{S.~Afanasiev} \affiliation{\jinrdubna}
\author{C.~Aidala} \affiliation{\columbia} \affiliation{\michigan}
\author{N.N.~Ajitanand} \affiliation{\stonybrkc}
\author{Y.~Akiba} \affiliation{\kek} \affiliation{\riken} \affiliation{\rikjrbrc}
\author{A.~Al-Jamel} \affiliation{\nmsu}
\author{J.~Alexander} \affiliation{\stonybrkc}
\author{K.~Aoki} \affiliation{\kyoto} \affiliation{\riken}
\author{L.~Aphecetche} \affiliation{\subatech}
\author{R.~Armendariz} \affiliation{\nmsu}
% ...
\collaboration{PHENIX Collaboration} \noaffiliation

\date{\today}


\begin{abstract}

Measurements of the midrapidity transverse energy distribution, 
$d\Et/d\eta$, are presented for $p$$+$$p$, $d$$+$Au, and Au$+$Au collisions 
at $\sqrt{s_{_{NN}}}=200$~GeV and additionally for Au$+$Au collisions at 
$\sqrt{s_{_{NN}}}=62.4$ and 130 GeV. The $d\Et/d\eta$ distributions are 
first compared with the number of nucleon participants $N_{\rm part}$, 
number of binary collisions $N_{\rm coll}$, and number of 
constituent-quark participants $N_{qp}$ calculated from a Glauber model 
based on the nuclear geometry. For Au$+$Au, $\mean{ d\Et/d\eta}/N_{\rm 
part}$ increases with $N_{\rm part}$, while $\mean{ d\Et/d\eta}/N_{qp}$ is 
approximately constant for all three energies. This indicates that the two 
component ansatz, 
$dE_{T}/d\eta \propto (1-x) N_{\rm part}/2 + x N_{\rm coll}$, 
which has been used to represent $E_T$ distributions, is simply a proxy 
for $N_{qp}$, and that the $N_{\rm coll}$ term does not represent a 
hard-scattering component in $E_T$ distributions. The $dE_{T}/d\eta$ 
distributions of Au$+$Au and $d$$+$Au are then calculated from the measured 
$p$$+$$p$ $E_T$ distribution using two models that both reproduce the 
Au$+$Au data.  However, while the number-of-constituent-quark-participant 
model agrees well with the $d$$+$Au data, the additive-quark model does 
not.

\end{abstract}

	
\maketitle

\section{Introduction}

\label{sec:introduction}

Measurements of midrapidity transverse energy distributions 
$d\Et/d\eta$ in $p$$+$$p$, $d$$+$Au and Au$+$Au collisions at \sqsn=200 GeV 
and Au$+$Au collisions at \sqsn=62.4 and 130 GeV are presented. The 
transverse energy \Et is a multiparticle variable defined as the sum
\begin{equation}
   \Et=\sum_i E_i\,\sin\theta_i \\ 
\label{eq:ETdef}
\end{equation}
\begin{equation}
   d\Et(\eta)/d\eta=\sin\theta(\eta)\, dE(\eta)/d\eta, 
\end{equation}

where $\theta$ is the polar angle, $\eta=-\ln \tan\theta/2$ is the 
pseudorapidity, $E_i$ is by convention taken as the kinetic energy for 
baryons, the kinetic energy + 2 $m_N$ for antibaryons, and the total 
energy for all other particles, and the sum is taken over all particles 
emitted into a fixed solid angle for each event.  In the present 
measurement as in previous measurements~\cite{Adcox:2001ry,Adler:2004zn} 
the raw \Et, denoted \Etemc, is measured in five sectors of the PHENIX 
lead-scintillator (PbSc) electromagnetic calorimeter 
(EMCal)~\cite{Adcox:2001ry} which cover the solid angle $|\eta|\leq 0.38$, 
$\Delta\phi=90^\circ+22.5^\circ$, and is corrected to total hadronic \Et, 
more properly $d\Et/d\eta|_{\eta=0}$, within a reference acceptance of 
$\Delta\eta=1.0, \Delta\phi=2\pi$ (details are given in section 
\ref{sec:analysis}).

The significance of systematic measurements of midrapidity $d\Et/d\eta$ 
and the closely related charged particle multiplicity distributions, 
$d\Nch/d\eta$, as a function of $A$ and $B$ in $A$+$B$ collisions is that 
they provide excellent characterization of the nuclear geometry of the 
reaction on an event-by-event basis, and are sensitive to the underlying 
reaction dynamics, which is the fundamental element of particle emission 
in $p$$+$$p$ and $A$+$B$ collisions at a given \sqsn . For instance, 
measurements of $d\Nch/d\eta$ in Au$+$Au collisions at the Relativistic 
Heavy Ion Collider (RHIC), as a function of centrality expressed as the 
number of participating nucleons, \Npart, do not depend linearly on \Npart 
but have a nonlinear increase of $\mean{d\Nch/d\eta}$ with increasing 
\Npart. The nonlinearity has been explained by a two component 
model~\cite{WangGyulassyPRL86,KharzeevNardiPLB507} proportional to a 
linear combination of \Ncoll and \Npart, with the implication that the 
\Ncoll term represents a contribution from hard scattering.  
Alternatively, it has been proposed that $d\Nch/d\eta$ is linearly 
proportional to the number of constituent-quark participants (NQP) 
model~\cite{EreminVoloshinPRC67}, without need to introduce a 
hard-scattering component. For symmetric systems, the NQP model is 
identical to the Additive Quark Model (AQM)~\cite{AQMPRD25} used in the 
1980's, to explain the similar nonproportionality of $d\Et/d\eta$ with 
\Npart in $\alpha-\alpha$ compared to $p$$+$$p$ collisions at \sqsn=31 
GeV~\cite{OchiaiZPC35}. In the AQM, constituent-quark participants in the 
two colliding nuclei are connected by color-strings; but with the 
restriction that only one color-string can be attached to a 
quark-participant. At midrapidity, the transverse energy production is 
proportional to the number of color-strings spanning between the 
projectile and the target nuclei. For asymmetric systems, such as 
$d$$+$Au, the models differ because the number of color-strings is 
proportional only to the number of quark-participants in the projectile 
(the lighter nucleus).  For symmetric A+A collisions, the number of 
quark-participants in the target is the same as number of 
quark-participants in the projectile, so the AQM is equivalent to the NQP 
model. These models will be described in detail and tested with the 
present data.

\section{Previous Measurements---A Historical Perspective}
\label{sec:historical}

\subsection{Charged Multiplicity Distributions}

The charged particle multiplicity or multiplicity density in rapidity, 
$d\Nch/dy$, is one of the earliest descriptive variables in high energy 
particle and nuclear physics dating from cosmic-ray studies~\cite{ET181}.  
An important regularity first observed in cosmic rays was that the 
produced pions have limited transverse momentum with respect to the 
collision axis, exponentially decreasing as $e^{-6p_T}$, commonly known as 
the ``Cocconi Formula''~\cite{Cocconi1961,Orear1964}.

By the early 1970's the framework for the study of this ``soft'' 
multi-particle physics was well in place. One of the important conceptual 
breakthroughs was the realization that the distribution 
of multiplicity for multiple particle production would not be Poisson 
unless the particles were emitted independently, without any correlation, 
but that short-range rapidity correlations were expected as a consequence 
of ``Regge-Pole-dominated'' reactions~\cite{MuellerPRD4}. In fact, in 
marked deviation from Poisson behavior, the total charged particle 
multiplicity distributions appeared to exhibit a universal form, ``KNO 
scaling''~\cite{KNONPB40} (or ``scaling in the mean''), when ``scaled'' at 
each \sqs by the average multiplicity---i.e. $d\Nch/dz$ was a universal 
function of the scaled multiplicity, $z\equiv \Nch/\mean{\Nch}$, where 
$\mean{\Nch}$ is the mean multiplicity at a given 
$\sqrt{s}$~\cite{UA1PLB107}.  In the mid 1980's, the UA5 group at the CERN 
Super Proton Synchrotron collider discovered that KNO scaling did not hold 
in general~\cite{UA5}, and found that their measured multiplicity 
distributions, both in limited rapidity intervals and over all phase space 
were described by negative binomial distributions (NBD), which since then 
have been shown to provide accurate descriptions for ${\Nch}$ 
distributions from high energy collisions of both particles and nuclei.

Also in this period, the central plateau of the rapidity distribution of 
identified charged particles, $d\Nch(y)/dy$, was discovered at the 
CERN-ISR~\cite{egJacob84-13}. Along with this discovery came the first 
interest to measure the multiplicity distribution in a restricted 
pseudorapidity range, $|\eta|\leq 1.5$, ``wide enough to allow for good 
statistics, yet sufficiently remote from the edge of the rapidity plateau 
to permit specific analysis of the central region''~\cite{Thome77}. The 
first suggestion to use multiplicity distributions in restricted regions 
of rapidity for the study of reaction dynamics, specifically quantum 
optical coherence effects in $p$$+$$p$ collisions, was made by Fowler and 
Weiner~\cite{FowlerWeiner77}, who emphasized the importance of using 
small-regions, where energy-momentum-conservation constraints would not be 
significant.

\subsection{\Et Distributions}

The phenomenology of \Et measurements, which evolved over a similar time 
period as that of multiplicity distributions, was based initially on the 
search for the jets of hard-scattering in $p$$+$$p$ collisions in 
``$4\pi$-hadron calorimeters'' as first proposed by 
Willis~\cite{WillisISAproc72} and then by Bjorken~\cite{BjorkenPRD8}, who 
specifically emphasized the need for the capability of measuring the total 
amount of energy emerging into small elements of solid angle to 
observe the event structure of what he called local cores (now jets) 
predicted for hard-scattering. Ochs and 
Stodolsky~\cite{OchsStodolskyPLB69} later proposed the veto of energy by a 
calorimeter in the forward direction, which was elaborated by Landshoff 
and Polkinghorne~\cite{LandshoffPolkPRD18} who coined the name 
``transverse energy'':  ``The energy not observed in the forward direction 
due to hard-scattering processes would be emitted as `transverse energy' 
''. The first experiment to measure an ``\Et distribution'' corresponding 
to the terminology used at present was the NA5 experiment at 
CERN~\cite{NA5PLB112}, which utilized a full azimuth hadronic calorimeter 
covering the region $-0.88<\eta<0.67$. They demonstrated that instead of 
finding jets~\cite{NA5PLB112}, ``The large \Et observed is the result of a 
large number of particles with a rather small transverse momentum.'' The 
close relationship between \Et and multiplicity distributions was shown in 
a measurement by UA1 in $\bar{\rm p}$+p collisions at \sqs=540 GeV at the 
CERN Super Proton Synchrotron collider~\cite{UA1-EP82-122}, with a full 
azimuth ``hermetic 
calorimeter'' covering $|\Delta\eta|\leq3$, which demonstrated that the 
``\Et measured in the calorimeter was strongly correlated to the measured 
multiplicity'' and that the KNO scaled \Et and \Nch distributions were 
``strikingly similar''. Ironically, this was to be presented at the same 
meeting (ICHEP82) at which UA2 presented the discovery of 
dijets~\cite{UA2JetICHEP82} in the region of a break in the steep 
exponential slope of an \Et distribution, to a flatter slope, 5--6 orders 
of magnitude down in cross section. Since then, it has been established 
that \Et and \Nch distributions are much less sensitive to hard-scattering 
than single inclusive measurements; and these distributions have been used 
to study the ``soft'' physics that dominates the $p$$+$$p$ inelastic cross 
section ~\cite{UA2-hard-soft-PLB165}. In fact, just a year after ICHEP82, 
Bjorken~\cite{BjorkenPRD27} stressed the importance of the region of the 
``central plateau'' of rapidity for the study of the evolution of the 
Quark Gluon Plasma and proposed $d\Et/dy|_{y=0}$ as an estimate of the 
co-moving energy density in a longitudinal expansion, proportional to the 
energy density in space, called the Bjorken Energy Density:
\begin{equation}
\epsilon_{Bj}= \frac{d\Et}{dy}  \frac{1}{\tau_0 \pi R^2}
\label{eq:epsilonBj}
\end{equation}
where $\tau_0$, the formation time, is usually taken as 1 fm/$c$ and $\pi 
R^2$ is the effective area of the collision. This formula is derived under 
the assumption that $\mean{\Et}$ per particle $\propto T$ for a thermal 
medium, which has nothing to do with hard scattering.

\subsection{Collisions of Relativistic Nuclei-Extreme Independent Models}
     
The first experiments specifically designed to measure the dependence of 
the charged particle multiplicity in high energy $p+A$ collisions as a 
function of the nuclear size were performed by Busza and 
collaborators~\cite{BuszaPRL34} at Fermilab using beams of 
$\sim$50--200~GeV/$c$ hadrons colliding with various fixed nuclear 
targets. They found the extraordinary result~\cite{BuszaPRL34} that the 
charged particle multiplicity density, $d\Nch/d\eta$, observed in 
proton+nucleus ($p$+A) interactions was not simply proportional to the 
number of collisions, but increased much more slowly. The other striking 
observation~\cite{HalliwellPRL39} was that a relativistic incident proton 
could pass through e.g. $\nu=4$ absorption-mean-free-paths of a target 
nucleus and emerge from the other side, and furthermore there was no 
intranuclear cascade of produced particles (a stark difference from what 
would happen to the same proton in a macroscopic 4 mean-free-path hadron 
calorimeter). In the forward fragmentation region of 200 GeV/$c$ $p$+$A$ 
collisions, within one unit of rapidity from the beam $y^{\rm beam}=6.0$, 
there was essentially no change in $d\Nch/d\eta$ as a function of $A$, 
while the peak of the distribution moved backwards from midrapidity 
(${y^{\rm cm}_{_{NN}}}\sim 3.0$) with increasing $A$ and the total 
multiplicity increased, resulting in a huge relative increase of 
multiplicity in the target fragmentation region, $\eta<1$ in the 
laboratory system.
    
These striking features of the $\sim 200$ GeV/$c$ fixed target 
hadron-nucleus data ($\sqsn\sim 19.4$ GeV) showed the importance of taking 
into account the time and distance scales of the soft multi-particle 
production process including quantum mechanical 
effects~\cite{FishbaneTrefilPRD9,FishbaneTrefilPLB51,GottfriedPRL32,ASGoldhPRD7,BialasCzyzPLB51,BoIngvarNPB88}. 
The observations had clearly shown that the target nucleus was rather 
transparent, so that a relativistic incident nucleon could make many 
successive collisions while passing through the nucleus and emerge intact. 
Immediately after a relativistic nucleon interacts inside a nucleus, the 
only thing that can happen consistent with relativity and quantum 
mechanics is for it to become an excited nucleon with roughly the same 
energy and reduced longitudinal momentum and rapidity. The relativistic 
nucleon remains in that state inside the nucleus, because the uncertainty 
principle and time dilation prevent it from fragmenting into particles 
until it is well outside the nucleus. This feature immediately eliminates 
the possibility of a cascade in the nucleus from the {rescattering} of the 
secondary products. Making the further assumptions (1) that an excited 
nucleon interacts with the same cross section as an unexcited nucleon and 
(2) that the successive collisions of the excited nucleon do not affect the 
excited state or its eventual fragmentation products~\cite{midFrankel}, 
leads to the conclusion 
that the elementary process for particle production in nuclear collisions 
is the excited nucleon.  This also leads to the prediction that the 
multiplicity in nuclear interactions should be proportional to the total 
number of projectile and target participants, rather than to the total 
number of collisions, which is called the wounded-nucleon model 
(WNM)~\cite{WNM}.  Common usage is to refer to the wounded nucleons (WN) 
as participants.

Interestingly, at midrapidity, the WNM works well only at roughly 
\sqsn$\sim 20$ GeV where it was discovered.  For \sqsn$\lsim$~5.4~GeV, 
particle production is smaller than the WNM due to the large 
stopping~\cite{E866Akiba} with reduced transparency; and the \Et 
distributions in $A$+$B$ collisions can be represented by sums of 
convolutions of the $p$+$A$ distribution according to the relative 
probability of the number of projectile participants, the 
wounded-projectile-nucleon model 
(WPNM)~\cite{FtPLB188,E802PLB197,E802PRC63}. For \sqsn$\geq 31$ GeV, 
particle production is larger than the WNM~\cite{BCMOR-alfalfa, AFSET89} 
and the AQM~\cite{AQMPRD25,OchiaiZPC35}, which is equivalent to a 
wounded-projectile-quark (color-string) model, has been used successfully. 
All three of the above models, as well as the models to be described 
below, are of the type referred to as extreme independent models (EIM). 
The effect of the nuclear geometry of the interaction can be calculated in 
EIMs, independently of the dynamics of particle production, which can be 
derived from experimental measurements, usually the $p$$+$$p$ (or $p$+$A$) 
measurement in the same detector. In fact, the first published 
measurements at the CERN~\cite{NA35PLB184} and BNL~\cite{E802ZPC38} fixed 
target heavy ion programs in 1986-87 were \Et distributions in which EIM, 
rather than cuts on centrality, were used to understand the data.

At RHIC (\sqsn$=19.6-200$ GeV), PHOBOS~\cite{Alver-Nch-PRC83} has shown 
that the WNM works in Au$+$Au collisions for the total multiplicity, \Nch, 
over the range $|\eta|<5.4$, while at midrapidity, the WNM fails---the 
multiplicity density per participant pair, 
$\mean{d\Nch/d\eta}/(\Npart/2)$, increases with increasing number of 
participants, in agreement with previous PHENIX 
results~\cite{Adcox:2000sp,Adcox:2001ry,Adler:2004zn}. Additionally, it 
has been shown using PHOBOS Au$+$Au 
data~\cite{EreminVoloshinPRC67,NouicerEPJC49} and discussed for other 
data~\cite{DeBhattPRC71} that the midrapidity $\mean{d\Nch/d\eta}$ as a 
function of centrality in Au$+$Au collisions is linearly proportional to 
the NQP model; however for symmetric systems this cannot be distinguished 
from the number of color-strings, the AQM~\cite{Bialas2008}. The present 
work completes the cycle and demonstrates, using midrapidity \Et 
distributions at \sqsn=200 GeV in the asymmetric $d$$+$Au system, as well 
as $p$$+$$p$ and Au$+$Au collisions, that the asymmetric $d$$+$Au 
measurement, which is crucial in distinguishing the color-string AQM from 
NQP models, clearly rejects the AQM and agrees very well with the NQP model.

While the concept of nucleon participants in collisions of nuclei is 
straightforward to understand, the concept of constituent-quark 
participants needs some elaboration. The nonrelativistic constituent-quark 
model~\cite{MGM64,Zweig} is the basis of understanding the observed 
spectrum of the meson and baryon elementary particles as bound states, 
i.e. $(q\bar{q})$ for mesons and $(qqq)$ for baryons. In addition to the 
masses and quantum numbers, other static properties such as the magnetic 
moments of baryons are also predicted in this model (see 
Refs.~\cite{Morpurgo,Kokkedee}, and references therein). However, these 
constituent-quarks are not the nearly massless $u$ and $d$ quarks 
(partons), called ``current quarks'' from their role in the currents of 
electroweak and QCD quantum field theories. The constituent-quarks are 
assumed to be complex objects or quasi-particles~\cite{ShuryakNPB116} made 
out of the point-like partons of QCD hard-scattering, the (current) 
quarks, anti-quarks and gluons. The constituent or valence quarks (valons) 
thus acquire masses on the order of 1/3 the nucleon mass (or 1/2 the 
$\rho$-meson mass), called ``chiral symmetry 
breaking''~\cite{Diakonov,WeinbergEPJC34}, when bound in the nucleon (or 
meson). According to Shuryak~\cite{ShuryakNPB116} (see also 
Ref.~\cite{AnisovichNPB133}), there are two scales for hadrons predicted 
in QCD, the confinement length given by the radii of hadrons, $R_{\rm 
conf}\approx 1$ fm $\approx R_{\rm hadron}$, as well as objects at the 
scale 1/3 smaller, the constituent-quarks (valons~\cite{HwaPRD22}). For 
instance, the consideration of constituent-quarks as `little bags' with 
application to the $\sigma_L/\sigma_T$ puzzle in deep inelastic 
lepton-hadron scattering and other hard processes was made by 
T.~Akiba~\cite{TAkibaPL109B}.

One other key feature of the constituent-quark model is additivity: the 
properties of a hadron are described as the independent sum of 
contributions of the individual quarks. In other words the three 
constituent-quarks in each nucleon in a nucleon-nucleon collision act like 
the three nucleons in each triton in a $^3$H+$^3$H collision: i.e. apart 
from their spatial correlation, the three nucleons in each triton act 
independently in the collision. This additive quark 
assumption~\cite{LevinJETPLett2,LipkinPRL16,AnisovichNPB133} gives the 
relation that the pion-nucleon total cross section is 2/3 the 
nucleon-nucleon total cross section, i.e. $\sigma (\pi p)/ \sigma (p 
p)=2/3$. The constituent-quark participant (NQP) model is simply an 
extension of this idea to multiplicity and $E_T$ distributions (``soft'' 
multi-particle physics) in $p$$+$$p$, $p$+A and A+A collisions. Although 
proposed first~\cite{AQMPRD25}, the AQM is a special case of the NQP model 
in which a color string connects two constituent-quarks which have 
scattered, and breaking of the color-string produces particles at 
midrapidity. However, in the AQM~\cite{AQMPRD25,OchiaiZPC35}, it is 
further assumed that multiple strings attached to the same projectile 
quark in a $p$+A collision coalesce and collapse into one color string, so 
that the AQM is effectively a wounded projectile quark model.

In this paper, we compare extreme-independent models of soft 
multi-particle production based on the number of fundamental elements 
taken as nucleon participants, nuclear collisions, constituent quarks and 
color-strings (AQM) with our measurement of transverse energy production. 
It will be shown that the ansatz, \mbox{$d\Et/d\eta\propto (1-x) \Npart/2 
+ x \Ncoll$}, does not imply that there is a hard scattering component in 
multi-particle production, consistent with the direct observations noted 
above. Thus, possible models motivated by the fact that half of the 
momentum of a nucleon is carried by gluons when probed at high $Q^2$ in 
hard-scattering are not considered and we limit our comparison to the 
nucleon and constituent-quark participant models of soft-multiparticle 
production widely used since the 1970's as discussed in the introduction.

\section{The PHENIX Detector}
\label{sec:detector}

\begin{figure}[htb] 
    \includegraphics[width=1.0\linewidth]{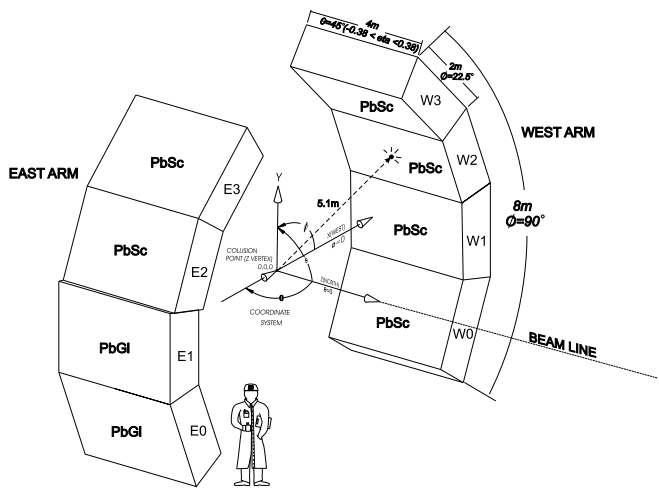}  
\caption{\label{fig:phenix2004}
Schematic diagram showing the locations of the PHENIX electromagnetic 
calorimeter sectors in the central arm spectrometer. The sectors labeled 
W1,W2,W3,E2 and E3 were used in this analysis}
\end{figure}

The PHENIX detector at Brookhaven National Laboratory's RHIC comprises two 
central spectrometer arms and two muon spectrometer arms. A comprehensive 
description of the detector components and performance can be found 
elsewhere~\cite{Adcox:2003zm}. The analysis described here utilizes five 
of the PbSc EMCal sectors~\cite{Adcox:2003zm} in the central arm 
spectrometers, as 
illustrated schematically in Figure \ref{fig:phenix2004}.  Each 
calorimeter sector covers a rapidity range of $|\eta| < 0.38$ and subtends 
$22.5^{o}$ in azimuth for a total azimuthal coverage of $112.5^{o}$. Each 
sector, whose front face is 5.1 m from the beam axis, is comprised of 
2,592 PbSc towers assembled in a 36 x 72 array. Each tower has a 5.535 cm 
x 5.535 cm surface area and an active depth of 37.5 cm corresponding to 
0.85 nuclear interaction lengths or 18 radiation lengths. The PbSc EMCal 
energy resolution for test beam electrons is $\frac{\Delta E}{E} = 
\frac{8.1\%}{\sqrt{E} (GeV)} \oplus$ 2.1\%, with a measured response 
proportional to the incident electron energy to within $ \pm 2\%$ over the 
range $0.3 \leq E_{e} \leq 40$ GeV.

A minimum-bias (MB) trigger for Au$+$Au, $d$$+$Au, and $p$$+$$p$ 
collisions is provided by two identical beam-beam counters (BBC), labeled 
North and South, each consisting of 64 individual \v{C}erenkov counters. 
The BBCs cover the full azimuthal angle in the pseudorapidity range 
$3.0<|\eta|<3.9$~\cite{Allen:2003zt}. For $p$$+$$p$ and $d$$+$Au 
collisions, events are required to have at least one counter fire in both 
the North and South BBCs. For Au$+$Au collisions, at least two counters 
must fire in both BBCs. Timing information from the BBCs are used to 
reconstruct the event vertex with a resolution of 6 mm for central Au$+$Au 
collisions.  All events are required to have an event vertex within 20 cm 
of the origin. Centrality determination in 200 GeV and 130 GeV Au$+$Au 
collisions~\cite{Adcox:2003zm} is based upon the total charge deposited in 
the BBCs and the total energy deposited in the Zero Degree Calorimeters 
(ZDC)~\cite{Allen:2003zt}, which are hadronic calorimeters covering the 
pseudorapidity range $|\eta| > 6$.  For 62.4 GeV Au$+$Au collisions, only 
the BBCs are used to determine centrality due to the reduced acceptance of 
the ZDCs at lower energies~\cite{Adare:2012wf}.

Table~\ref{tab:datasets} gives a summary of 
the 2003 and 2004 data sets used in this analysis.
Previously, PHENIX has studied transverse energy production in Au$+$Au 
collisions at \sqsn=200 GeV, 130 GeV, and 19.6 
GeV~\cite{Adcox:2001ry,Adler:2004zn} and shown that for \Et measurements 
at midrapidity at a collider the EMCal acts as a thin but effective hadron 
calorimeter.  Presented here is an extended analysis of 200 GeV Au$+$Au 
collision data taken during 2004 with the magnetic field turned on that 
increases the statistics of the previous analysis by a factor of 494 with 
132.9 million MB events. These new results are consistent 
with the previously published results~\cite{Adcox:2001ry,Adler:2004zn}.

The average luminosity delivered by RHIC has improved dramatically each 
year, by a factor of 5.75 for $p$$+$$p$ collisions and 4.5 for $d$$+$Au 
collisions from the 2003 to 2008 running periods. Due to the readout 
electronics implemented for the EMCal, with a pile-up window of 428
nsec, the increased luminosity results in an increasing rate-dependent 
background in the minimum-bias event sample due to multiple collisions, or 
pile-up, that artificially raises the transverse energy recorded in an 
event.  To minimize this background, the 200 GeV $p$$+$$p$ and 200 
GeV $d$$+$Au data samples presented hear are from the earlier 2003 running 
period. 

\begin{table}[htb]
\caption{\label{tab:datasets}
Summary of the data sets used in this analysis. $N_{\rm events}$ 
represents the number of MB events analyzed and 
$\mathcal{L}_{\rm ave}$ represents the average RHIC luminosity for the 
dataset.}
\begin{ruledtabular}
\begin{tabular}{ccccc}
\sqsn (GeV) & System & Year & $N_{\rm events}$ 
& $\mathcal{L}_{\rm ave} (cm^{-2}s^{-1})$\\[0.2pc]\hline
200 & Au$+$Au & 2004 & 132.9 M & $5 \times 10^{26}$\\
62.4 & Au$+$Au & 2004 & 20.0 M & $0.6 \times 10^{26}$\\
200 & $d$$+$Au & 2003 & 50.1 M & $3 \times 10^{28}$\\
200 & $p$$+$$p$ & 2003 & 14.6 M & $4 \times 10^{30}$\\
\end{tabular}
\end{ruledtabular}
\end{table}

\section{Data Analysis}
\label{sec:analysis}

The analysis procedure for $d\Et/d\eta$ is described in detail 
in~\cite{Adler:2004zn} and summarized here. The absolute energy scale for 
the PbSc EMCal was calibrated using the $\pi^{0}$ mass peak from pairs of 
reconstructed EMCal clusters. The uncertainty in the absolute energy scale 
is 2\% in the 62.4 GeV Au$+$Au dataset and 1.5\% in the 200 GeV Au$+$Au, 
$p$$+$$p$, and $d$$+$Au datasets. The transverse energy for each event was 
computed using clusters with an energy greater than 30 MeV composed of 
adjacent towers each with a deposited energy of more than 10 MeV. Faulty 
towers and all towers in a $3x3$ tower area around any faulty tower are 
excluded from the analysis.

\begin{figure}[htb] 
\includegraphics[width=1.0\linewidth]{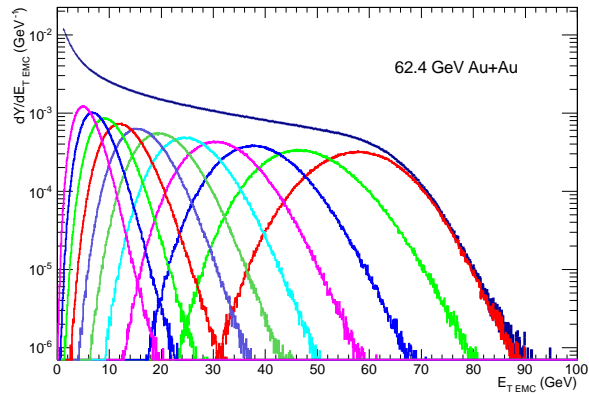}
\caption{\label{fig:auau62Overlay}(Color online) 
\Etemc distributions for $\sqrt{s_{\rm NN}}=62.4$ GeV Au$+$Au collisions. 
Shown are the MB distribution along with the distributions in 
5\% wide centrality bins selected using the BBCs.  All the plots are 
normalized so that the integral of the MB distribution is 
unity.}
\end{figure}

The raw spectra of the measured transverse energy \Etemc in the fiducial 
aperture are given as histograms of the number of entries in a given raw 
\Etemc bin such that the total number of entries sums up to the number of 
BBC counts. The distributions are then normalized to integrate to unity. 
As an example, the \Etemc distributions as a function of centrality in 5\% 
wide centrality bins are shown in Fig.~\ref{fig:auau62Overlay} for 62.4 
GeV Au$+$Au collisions.

\begin{figure}[htb] 
\includegraphics[width=1.0\linewidth]{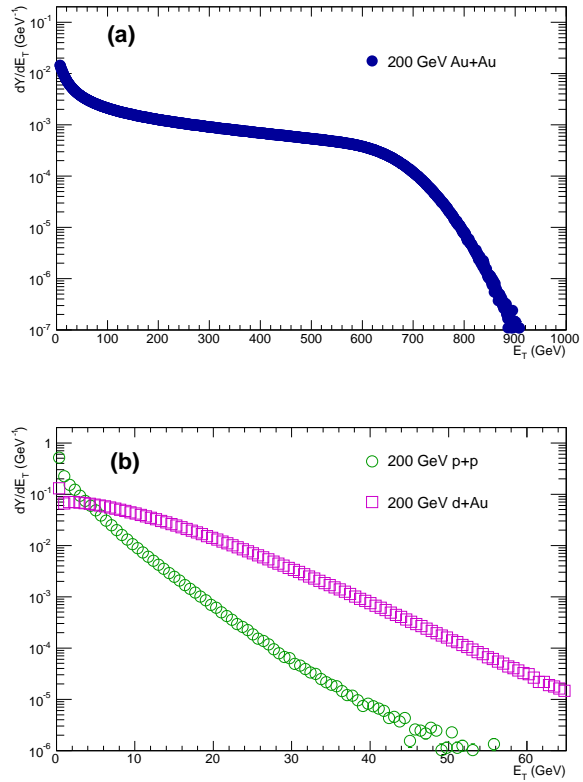} 
\caption{\label{fig:f1}(Color online) 
Corrected $\Et=d\Et/d\eta|_{\eta=0}$ distributions at $\sqrt{s_{\rm 
NN}}=200$ GeV for 5 sectors of PbSc (a) Au$+$Au; (b) $p$$+$$p$, 
$d$$+$Au. The correction factors for each dataset are listed in Table 
\ref{tab:corrections}. All the plots are normalized so that the integral 
of each distribution is unity.}
\end{figure}

To obtain the total hadronic \Et within a reference acceptance of 
$\Delta\eta=1.0, \Delta\phi=2\pi$, more properly $d\Et/d\eta|_{\eta=0}$, 
from the measured raw transverse energy, \Etemc, several corrections are 
applied. The total correction can be decomposed into three main 
components. First is a correction by a factor of 4.188 to account for the 
fiducial acceptance. Second is a correction by a factor of 1.262 for 200 
GeV Au$+$Au, 1.236 for 62.4 GeV Au$+$Au, 1.196 for 200 GeV $d$$+$Au, and 
1.227 for 200 GeV $p$$+$$p$ to account for disabled calorimeter towers not 
used in the analysis. Third is a factor, $k$, which is the ratio of the 
total hadronic \Et in the fiducial aperture to the measured \Etemc. 
Details on the estimate of the values of the $k$ factor are given below. 
The total correction scale factors are obtained by multiplying these three 
components and are listed in Table \ref{tab:corrections}. The corrected 
MB distributions for 200 GeV Au$+$Au, $d$$+$Au, and $p$$+$$p$ are 
shown in Fig~\ref{fig:f1}.

\begin{table}[htb]
\caption{\label{tab:corrections}
Summary of the total correction scale factors applied to the measured raw 
transverse energy, \Etemc, to obtain $d\Et/d\eta|_{\eta=0}$ for each 
dataset.}
\begin{ruledtabular}
\begin{tabular}{ccc}
\sqsn (GeV) & System & Correction Factor\\[0.2pc]\hline
200 & Au$+$Au & $6.87 \pm 0.40$\\
62.4 & Au$+$Au & $6.73 \pm 0.39$\\
200 & $d$$+$Au & $6.51 \pm 0.54$\\
200 & $p$$+$$p$ & $6.68 \pm 0.56$\\
\end{tabular}
\end{ruledtabular}
\end{table}

The $k$ factor comprises three components. The first component, denoted 
$k_{\rm response}$, is due to the fact that the EMCal was designed for the 
detection of electromagnetic particles~\cite{Adcox:2001ry}. Hadronic 
particles passing through the EMCal only deposit a fraction of their total 
energy. The average EMCal response is estimated for the various particle 
species using the HIJING event generator~\cite{Wang:1991hta} processed 
through a {\sc geant}-based Monte Carlo simulation of the PHENIX detector. 
The HIJING particle composition and $p_{T}$ spectra are adjusted to 
reproduce the identified charged particle spectra and yields measured by 
PHENIX. For all of the data sets, 75\% of the total energy incident on the 
EMCal is measured, thus $k_{\rm response}$ = 1/0.75 = 1.33. The second 
component of the $k$ factor, denoted $k_{inflow}$, is a correction for 
energy inflow from outside the fiducial aperture of the EMCal. This energy 
inflow arises from two sources: from parent particles with an original 
trajectory outside of the fiducial aperture whose decay products are 
incident within the fiducial aperture, and from particles that reflect off 
of the PHENIX magnet poles into the EMCal fiducial aperture. The energy 
inflow contribution is 24\% of the measured energy, thus $k_{\rm inflow}$ = 
1-0.24 = 0.76. The third component of the $k$ factor, denoted 
$k_{\rm losses}$, is due to energy losses. There are three components to the 
energy loss: from particles with an original trajectory inside the 
fiducial aperture of the EMCal whose decay products are outside of the 
fiducial aperture (10\%), from energy losses at the edges of the EMCal 
(6\%), and from energy losses due to the energy thresholds (6\%). The 
total contribution from energy losses is 22\%, thus $k_{losses}$ = 
1/(1-0.22) = 1.282. The total $k$ factor correction is $k = k_{\rm response} 
\times k_{\rm inflow} \times k_{\rm losses}$ = 1.30.

When plotting transverse energy production as a function of centrality, 
systematic uncertainties are decomposed into three types. Type A 
uncertainties are point-to-point uncertainties that are uncorrelated 
between bins and are normally added in quadrature to the statistical 
uncertainties. However, because there are no Type A uncertainties in this 
analysis, the vertical error bars represent statistical uncertainties 
only. Type B uncertainties are bin-correlated such that all points move in 
the same direction, but not necessarily by the same factor. These are 
represented by a pair of lines bounding each point. Type C uncertainties 
are normalization uncertainties in which all points move by the same 
factor independent of each bin.  These are represented as a single error 
band on the right hand side of each plot. In addition, there is an 
uncertainty on the estimate of the value of $<N_{\rm p0.0art}>$ at each 
centrality that is represented by horizontal error lines.

There are two contributions to Type B uncertainties, which are added in 
quadrature to obtain the total Type B uncertainty. The first contribution 
to Type B uncertainties arises from the uncertainty in the trigger 
efficiency.  The method by which the trigger efficiency is determined is 
described in~\cite{Adler:2004zn}. The BBC trigger efficiency is 
92.2\%$^{+2.5\%}_{-3.0\%}$ for 200 GeV and 130 GeV Au$+$Au collisions, 
83.7\%$\pm$3.2\% for 62.4 GeV Au$+$Au collisions, 88\%$\pm$4\% for 200 GeV 
$d$$+$Au collisions, and 54.8\%$\pm$5.3\% for 200 GeV $p$$+$$p$ 
collisions~\cite{Adler:2006wg}.  Because the centrality is defined for a 
given event as a percentage of the total geometrical cross section, an 
uncertainty in the trigger efficiency translates into an uncertainty in 
the centrality definition. This uncertainty is estimated by measuring the 
variation in $d\Et/d\eta$ by redefining the centrality using trigger 
efficiencies that vary by $\pm 1$ standard deviation. The second 
contribution to Type B uncertainties is the uncertainty due to random 
electronic noise in the EMCal towers. The noise, or background, 
contribution is estimated to be consistent with zero with uncertainties 
tabulated in Table \ref{tab:sysErrors} by measuring the average energy 
deposited per sector in events where all the particles are screened by the 
central magnet pole tips by requiring an interaction z-vertex of $+50 < z 
< +60$ cm and $-50 < z < -60$ cm. A summary of the magnitudes of the Type 
B uncertainty contributions is listed in Table \ref{tab:sysErrors}.

There are several components to Type C uncertainties, which are also added 
in quadrature to obtain the total Type C uncertainty. The first 
contribution is the uncertainty of the energy response estimate.  This 
uncertainty includes uncertainties in the absolute energy scale, 
uncertainties in the estimate of the hadronic response, and uncertainties 
from energy losses on the EMCal edges and from energy thresholds.  The 
uncertainties in the hadronic response estimate include a 3\% uncertainty 
estimated using a comparison of the simulated energy deposited by hadrons 
with different momenta with test beam data~\cite{Aphecetche:2003zr} along 
with an additional 1\% uncertainty in the particle composition and 
momentum distribution. Other Type C uncertainties include an uncertainty 
in the estimate of the EMCal acceptance, an uncertainty in the calculation 
of the fraction of the total energy incident on the EMCal fiducial area 
(losses and inflow), and an uncertainty in the centrality determination.  
A summary of the magnitudes of the Type C uncertainty contributions is 
listed in Table \ref{tab:sysErrors}. For the MB distributions, 
the uncertainties on the scale factors previously quoted contain only Type 
C uncertainties from the energy response, acceptance, and from losses and 
inflow.

\begin{table*}[htb]
\caption{
Summary of the systematic uncertainties given in percent. Listed are 
uncertainties classified as Type B and Type C. A range is given for Type B 
uncertainties with the first number corresponds to the most central bin 
and the second number corresponds to the most peripheral bin.}
\label{tab:sysErrors}
\begin{ruledtabular} \begin{tabular}{ccccccc}
  &   
& \multicolumn{3}{c}{Au$+$Au} 
& $d$$+$Au  
& $p$$+$$p$ \\
Error type & System 
& 200 GeV 
& 130 GeV 
& 62.4 GeV 
& 200 GeV 
& 200 GeV \\
\hline
C & Energy Resp. & 3.9\% & 3.8\% & 3.9\% & 3.9\% & 3.9\%\\
C & Acceptance & 3.0\% & 3.0\% & 3.0\% & 3.0\% & 3.0\%\\
C & Losses and Inflow & 3.0\% & 3.0\% & 3.0\% & 3.0\% & 3.0\%\\
C & Centrality & 0.5\% & 0.5\% & 0.5\% & n/a & n/a\\
B & Trigger & 0.3\%-16\% & 0.3\%-16\% & 0.44\%-16\% & n/a & n/a\\
B & Background & 0.2\%-6.0\% & 0.4\%-10.0\% & 0.375\%-13.3\% & 6.0\% & 6.0\%\\
\end{tabular} \end{ruledtabular}
\end{table*}

\begin{table*}[htb]
\caption{
The inelastic quark-quark cross sections used for each collision energy to 
reproduce the inelastic nucleon-nucleon cross section.}
\label{tab:qqCross}
\begin{ruledtabular}
\begin{tabular}{ccccc}
& \sqsn (GeV) & $\sigma^{\rm inel}_{NN}$ (mb) & $\sigma^{\rm inel}_{qq}$ (mb) & \\
\hline
& 200 & 42 & 9.36 & \\
& 130 & 40 & 8.60 & \\
& 62.4 & 35.6 & 7.08 & \\
\end{tabular}
\end{ruledtabular}
\end{table*}

\section{Estimating the Number of Nucleon and Quark Participants}
\label{sec:Glauber}

\begin{figure}[htb] 
\includegraphics[width=1.0\linewidth]{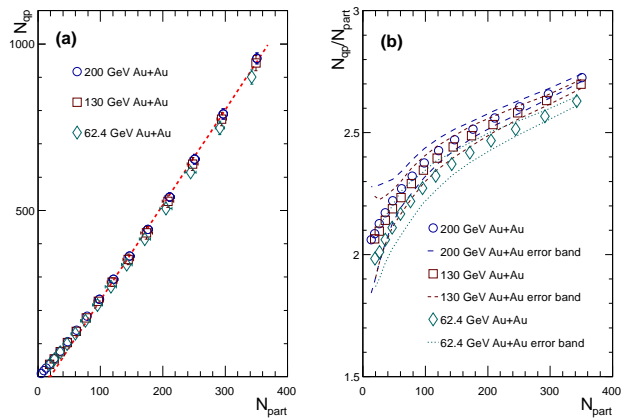} 
\caption{(Color online) 
(a) The number of quark participants as a function of the number of 
nucleon participants. The error bars represent the systematic uncertainty 
estimate on the MC-Glauber calculation. The dashed line is a linear fit to 
the 200 GeV Au$+$Au points with $N_{\rm part}>100$ to illustrate the 
nonlinearity of the correlation at low values of $N_{\rm part}$. 
(b) The ratio of the number of quark participants to the number of 
nucleon participants as a function of the number of nucleon participants. 
The error bands represent the systematic uncertainty estimate on the 
MC-Glauber calculation.}
    \label{fig:npartVsNq}
\end{figure}

A Monte-Carlo-Glauber (MC-Glauber) model calculation~\cite{Miller:2007ri} 
is used to obtain estimates for the number of nucleon participants at each 
centrality using the procedure described in~\cite{Adler:2004zn}. A similar 
procedure can be used to estimate the number of quark participants, 
$N_{qp}$, at each centrality. The quark-quark inelastic cross section for 
each collision energy is determined such that the inelastic 
nucleon-nucleon cross section is reproduced. The MC-Glauber calculation is 
then implemented so that the fundamental interactions are quark-quark 
rather than nucleon-nucleon collisions. Initially, the nuclei are 
assembled by distributing the centers of the nucleons according to a 
Woods-Saxon distribution. Once a nucleus is assembled, three quarks are 
then distributed around the center of each nucleon. The spatial 
distribution of the quarks is given by the Fourier transform of the form 
factor of the proton:
\begin{equation}
   \rho^{proton}(r) = \rho^{proton}_{0} \times e^{-ar},
\end{equation}
where $a = \sqrt{12}/r_{m} = 4.27$ fm$^{-1}$ and $r_{m}=0.81$ fm is the 
rms charge radius of the proton~\cite{HofstadterRMP}. The coordinates 
of the two colliding nuclei are shifted relative to each other by the 
impact parameter. A pair of quarks, one from each nucleus, interact with 
each other if their distance $d$ in the plane transverse to the beam axis 
satisfies the condition
\begin{equation}
   d < \sqrt{\frac{\sigma^{\rm inel}_{qq}}{\pi}},
\end{equation}
where $\sigma^{\rm inel}_{qq}$ is the inelastic quark-quark cross section, 
which is varied for the case of nucleon-nucleon collisions until the known 
inelastic nucleon-nucleon cross section is reproduced and then used for 
the A+A calculations. The resulting inelastic quark-quark cross sections 
are tabulated in Table \ref{tab:qqCross}. Figure \ref{fig:npartVsNq}a 
shows the number of quark participants as a function of the number of 
nucleon participants. The relationship is nonlinear, especially for low 
values of $N_{\rm part}$. Figure \ref{fig:npartVsNq}b shows the resulting 
ratio of the number of quark participants to the number of nucleon 
participants as a function of the number of nucleon participants.

\section{$d\Et/d\eta$ Results}
\label{sec:results}

\begin{figure}[htb] 
\includegraphics[width=1.0\linewidth]{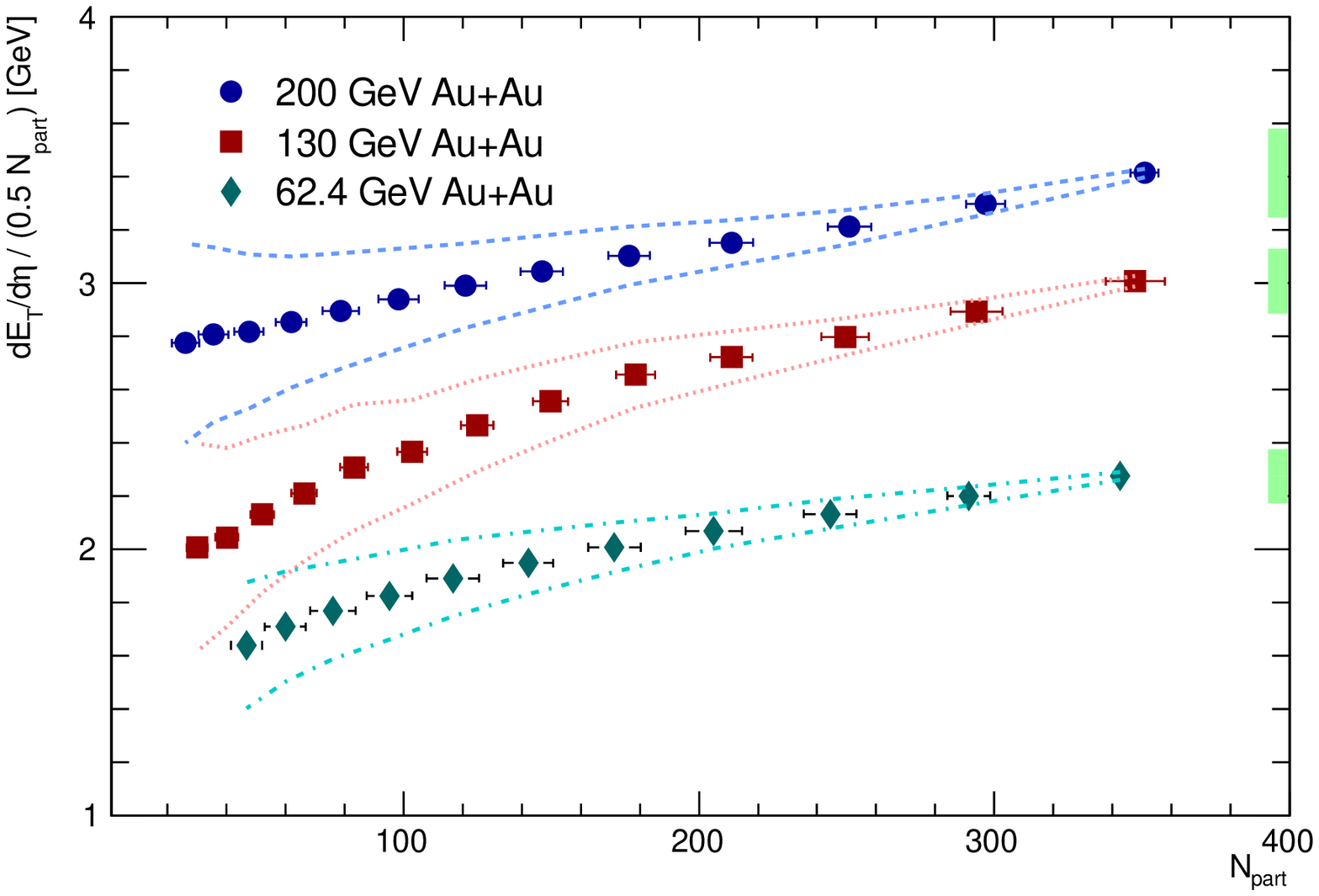} 
\caption{(Color online) 
$d\Et/d\eta$ normalized by the number of participant pairs as a function 
of the number of participants for Au$+$Au collisions at \sqsn=200, 130, 
and 62.4 GeV. The Type A uncertainties are represented by error bars about 
each point. The Type B uncertainties are represented by the lines bounding 
each point. The Type C uncertainties are represented by the error bands to 
the right of the most central data point.}
    \label{fig:detNormNpart}
\end{figure}

\begin{figure}[htb] 
\includegraphics[width=1.0\linewidth]{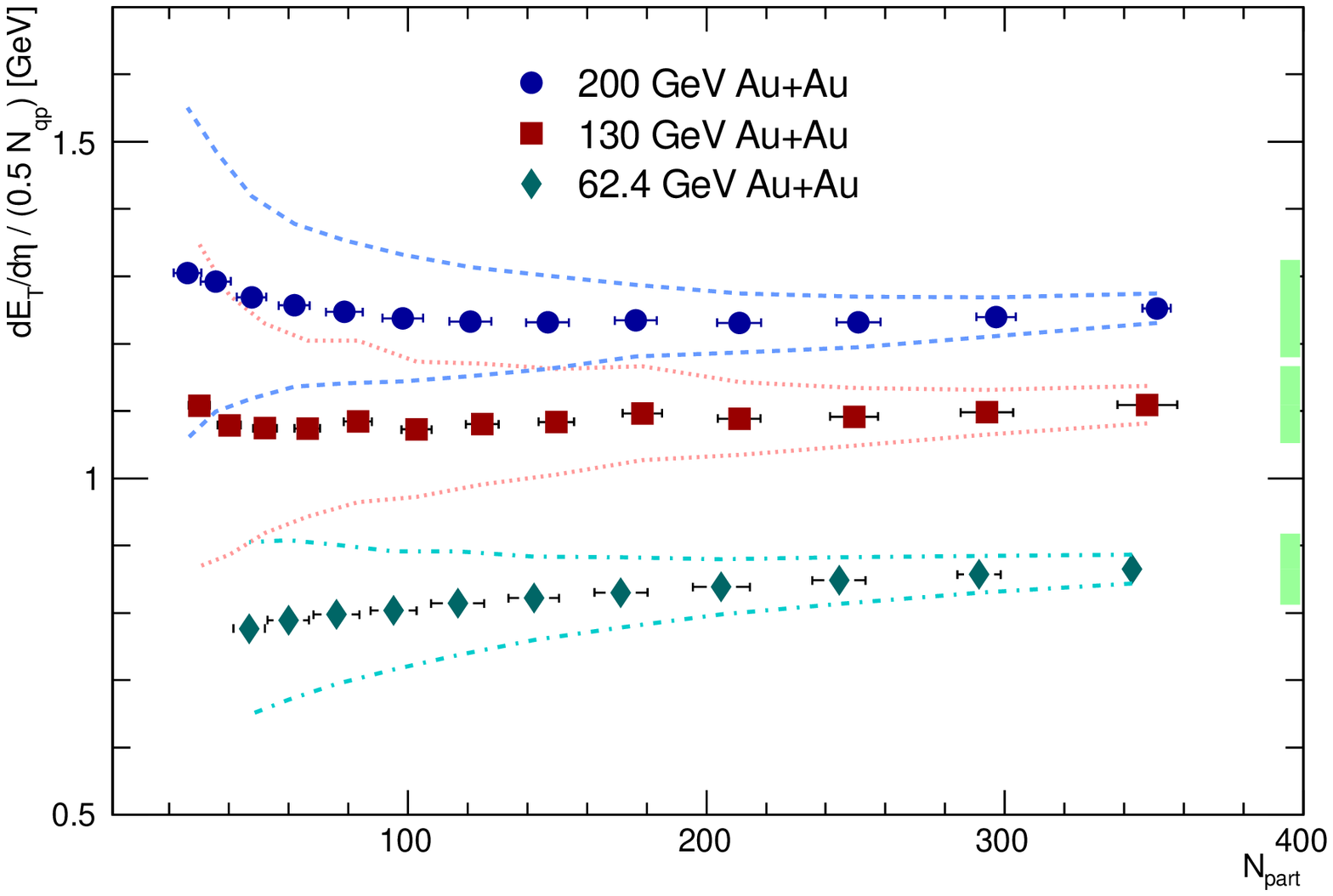} 
\caption{(Color online) 
$d\Et/d\eta$ normalized by the number of participant quark pairs as a 
function of the number of participants for Au$+$Au collisions at 
\sqsn=200, 130, and 62.4 GeV. The Type A uncertainties are represented by 
error bars about each point. The Type B uncertainties are represented by 
the lines bounding each point. The Type C uncertainties are represented by 
the error bands to the right of the most central data point.}
\label{fig:detNormNquark}
\end{figure}

\begin{figure}[htb] 
\includegraphics[width=1.0\linewidth]{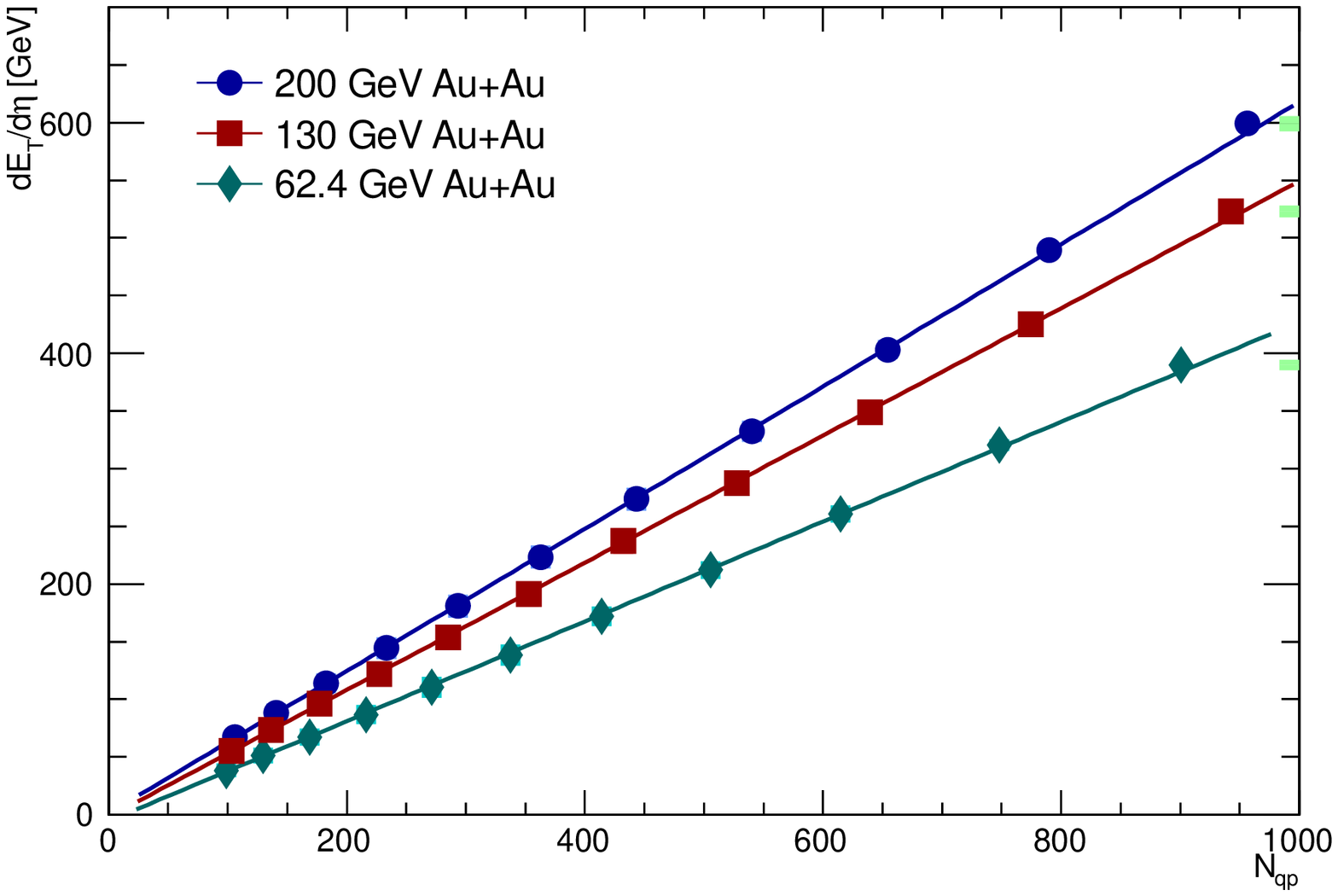} 
\caption{(Color online) 
$d\Et/d\eta$ as a function of the number of quark participants for Au$+$Au 
collisions at \sqsn=200, 130, and 62.4 GeV. The Type A uncertainties are 
represented by error bars about each point. The Type B uncertainties are 
represented by error bands about each point shown. The type A and type B 
uncertainties are typically less than the size of the data point. The Type 
C uncertainties are represented by the error bands to the right of the 
most central data point. The lines are straight line fits to the data.}
\label{fig:detdetaNquark}
\end{figure}

The distribution of $d\Et/d\eta$ normalized by the number of participant 
pairs as a function of the number of participants is shown in Figure 
\ref{fig:detNormNpart} for Au$+$Au collisions at \sqsn=200, 130, and 62.4 
GeV. The data are also tabulated in Table \ref{tab:auau200detdeta} for 200 
GeV Au$+$Au, Table \ref{tab:auau130detdeta} for 130 GeV Au$+$Au, and Table 
\ref{tab:auau062detdeta} for 62.4 GeV Au$+$Au collisions. For all 
collision energies, the increase seen as a function of \Npart is 
nonlinear, showing a saturation towards the more central collisions.  
However, when $d\Et/d\eta$ is normalized by the number of quark 
participant pairs, as shown in Figure \ref{fig:detNormNquark}, the data 
are consistently flat within the systematic uncertainties. Transverse 
energy production can also be plotted as a function of the number of quark 
participants as shown in Figure \ref{fig:detdetaNquark}.  The data for 
each collision energy are well described by a straight line as shown. The 
slope parameters for each collision energy are summarized in Table 
\ref{tab:detVsNq}. The consistency with zero of the values of the 
intercept $b$ establish a linear proportionality of \Et with \Nqp . To 
summarize, transverse energy production scales {\em linearly} with the 
number of constituent-quark participants, in contrast to the nonlinear 
relationship between transverse energy and the number of participating 
nucleons.

This nonlinear relationship has been successfully parametrized as a 
function of centrality~\cite{Adcox:2000sp,WangGyulassyPRL86,KharzeevNardiPLB507}:
 \begin{equation}
{d\Et^{\rm AA}/d\eta}=({d\Et^{pp}/d\eta})\ [(1-x)\,\mean{\Npart}/2 +x\,\mean{\Ncoll} ],
 \label{eq:crazy}
 \end{equation}
with the implication that the proportionality to \Ncoll is related to a 
contribution of hard-scattering to \Nch and \Et 
distributions~\cite{WangGyulassyPRL86,KharzeevNardiPLB507}. This seems to 
contradict the extensive measurements of \Nch and \Et distributions in 
$p$$+$$p$ collisions described in Sec.~\ref{sec:historical} which show that 
these distributions represent measurements of the ``soft'' multiparticle 
physics that dominates the $p$$+$$p$ inelastic cross section. Another 
argument against a hard-scattering component that the shape of the $ 
d\Nch/d\eta/(0.5 \Npart)$ vs. \Npart curves as in 
Fig.~\ref{fig:detNormNpart} is also the same at 2.76~TeV Pb$+$Pb 
collisions~\cite{ALICEPRL106} although the jet cross section increases by 
a very large factor. Furthermore, any supposed hard-component in the $p$$+$$p$ 
distributions would be suppressed in A+A collisions~\cite{Adcox:2001jp}. 
This apparent conflict can be resolved if Eq.~\ref{eq:crazy} is just a 
proxy for the correct description of the underlying physics, because
$d\Et^{\rm AA}/d\eta$ is strictly proportional to \Nqp 
(Fig.~\ref{fig:detdetaNquark}, Table~\ref{tab:detVsNq}). Using \Npart, 
\Ncoll and \Nqp as a function of centrality, with the value 
$x=0.08$~\cite{Adcox:2000sp,PHOBOSPRC70}, the ansatz in brackets in 
Eq.~\ref{eq:crazy} is compared to \Nqp as a function of centrality 
(Table~\ref{tab:auau200NcNpNqRatio}). The striking result is that the 
ratio $\Nqp/[(1-x)\,\mean{\Npart}/2 +x\,\mean{\Ncoll} ]=3.88$ on the 
average and varies by less than 1\% over the entire range except for the 
most peripheral bin where it drops by 5\%. This result demonstrates that 
rather than implying a hard-scattering component in \Nch and \Et 
distributions, Eq.~\ref{eq:crazy} is instead a proxy for the number of 
constituent-quark-participants \Nqp as a function of centrality.

It is important to point out that the relationship breaks down more 
seriously for $p$$+$$p$ collisions, with a ratio of 2.99 
(Table~\ref{tab:auau200NcNpNqRatio}). This is consistent with the 
PHOBOS~\cite{PHOBOSPRC70} result that a fit of Eq.~\ref{eq:crazy} to 
$\mean{d\Nch^{\rm AA}/d\eta}$ leaving $\mean{d\Nch^{pp}/d\eta}$ as a free 
parameter also projects above the $p$$+$$p$ measurement. Because the key to 
the utility of Extreme Independent Models is that the $p$$+$$p$ data, 
together with an independent calculation of the nuclear geometry can be 
used to predict the A+A distributions, we now turn to the analysis of the 
$p$$+$$p$, $d$$+$Au and Au$+$Au \Et distributions at \sqsn=200 GeV in terms of 
these models to see whether the extrapolation from the $p$$+$$p$ data using 
constituent-quark participants is more robust than from the ansatz.

\begin{table*}[hbt]
\caption{\label{tab:auau200detdeta}
Transverse energy production results for 200 GeV Au$+$Au collisions. 
Listed are the total uncertainties (Type A, Type B, and Type C) for each 
centrality bin.}
\begin{ruledtabular} \begin{tabular}{rccccc}
Centrality & $<N_{\rm part}>$ & $<N_{qp}>$ & $\frac{dE_{T}}{d\eta}$ [GeV] & $\frac{1}{0.5<N_{\rm part}>}$ $\frac{dE_{T}}{d\eta}$ [GeV] & $\frac{1}{0.5<N_{qp}>}$ $\frac{dE_{T}}{d\eta}$ [GeV] \\\hline
0\%--5\%   & 350.9 $\pm$ 4.7 & 956.6 $\pm$ 16.2 & 599.0 $\pm$ 34.7 & 3.41 $\pm$ 0.20 & 1.25 $\pm$ 0.08 \\
5\%--10\%  & 297.0 $\pm$ 6.6 & 789.8 $\pm$ 15.3 & 489.7 $\pm$ 28.9 & 3.29 $\pm$ 0.19 & 1.24 $\pm$ 0.08 \\
10\%--15\% & 251.0 $\pm$ 7.3 & 654.2 $\pm$ 14.5 & 403.0 $\pm$ 25.0 & 3.21 $\pm$ 0.19 & 1.23 $\pm$ 0.08 \\
15\%--20\% & 211.0 $\pm$ 7.3 & 540.2 $\pm$ 12.3 & 332.5 $\pm$ 21.2 & 3.15 $\pm$ 0.20 & 1.23 $\pm$ 0.08 \\
20\%--25\% & 176.3 $\pm$ 7.0 & 443.3 $\pm$ 10.4 & 273.6 $\pm$ 18.6 & 3.10 $\pm$ 0.21 & 1.23 $\pm$ 0.09 \\
25\%--30\% & 146.8 $\pm$ 7.1 & 362.8 $\pm$ 12.2 & 223.4 $\pm$ 16.4 & 3.04 $\pm$ 0.22 & 1.23 $\pm$ 0.09 \\
30\%--35\% & 120.9 $\pm$ 7.0 & 293.3 $\pm$ 11.0 & 180.8 $\pm$ 14.3 & 2.99 $\pm$ 0.23 & 1.23 $\pm$ 0.10 \\
35\%--40\% & 98.3  $\pm$ 6.8 & 233.5 $\pm$ 9.2  & 144.5 $\pm$ 12.6 & 2.94 $\pm$ 0.25 & 1.24 $\pm$ 0.11 \\
40\%--45\% & 78.7  $\pm$ 6.1 & 182.7 $\pm$ 6.8  & 113.9 $\pm$ 10.9 & 2.90 $\pm$ 0.27 & 1.25 $\pm$ 0.12 \\
45\%--50\% & 61.9  $\pm$ 5.2 & 140.5 $\pm$ 5.3  & 88.3  $\pm$ 9.3  & 2.85 $\pm$ 0.29 & 1.26 $\pm$ 0.14 \\
50\%--55\% & 47.6  $\pm$ 4.9 & 105.7 $\pm$ 5.5  & 67.1  $\pm$ 8.1  & 2.82 $\pm$ 0.33 & 1.27 $\pm$ 0.15 \\
55\%--60\% & 35.6  $\pm$ 5.1 & 77.3  $\pm$ 6.8  & 50.0  $\pm$ 6.7  & 2.81 $\pm$ 0.36 & 1.29 $\pm$ 0.17 \\
60\%--65\% & 26.1  $\pm$ 4.7 & 55.5  $\pm$ 7.1  & 36.2  $\pm$ 5.4  & 2.77 $\pm$ 0.40 & 1.30 $\pm$ 0.20 \\
\end{tabular} \end{ruledtabular}
\end{table*}

\begin{table*}[hbt]
\caption{\label{tab:auau130detdeta}
Transverse energy production results for 130 GeV Au$+$Au collisions. 
Listed are the total uncertainties (Type A, Type B, and Type C) for each 
centrality bin.}
\begin{ruledtabular} \begin{tabular}{rccccc}
Centrality & $<N_{\rm part}>$ & $<N_{qp}>$ & $\frac{dE_{T}}{d\eta}$ [GeV] & $\frac{1}{0.5<N_{\rm part}>}$ $\frac{dE_{T}}{d\eta}$ [GeV] & $\frac{1}{0.5<N_{qp}>}$ $\frac{dE_{T}}{d\eta}$ [GeV] \\
\hline
0\%--5\%   & 347.7 $\pm$ 10.0 & 942.6 $\pm$ 22.6 & 522.8 $\pm$ 27.7 & 3.01 $\pm$ 0.16 & 1.11 $\pm$ 0.06 \\
5\%--10\%  & 294.0 $\pm$ 8.9  & 774.7 $\pm$ 20.3 & 425.2 $\pm$ 23.3 & 2.89 $\pm$ 0.16 & 1.10 $\pm$ 0.07 \\
10\%--15\% & 249.5 $\pm$ 8.0  & 639.6 $\pm$ 19.4 & 349.0 $\pm$ 20.3 & 2.80 $\pm$ 0.16 & 1.09 $\pm$ 0.07 \\
15\%--20\% & 211.0 $\pm$ 7.2  & 527.7 $\pm$ 18.3 & 287.2 $\pm$ 18.3 & 2.72 $\pm$ 0.17 & 1.09 $\pm$ 0.08 \\
20\%--25\% & 178.6 $\pm$ 6.6  & 432.5 $\pm$ 19.0 & 237.1 $\pm$ 16.6 & 2.66 $\pm$ 0.19 & 1.10 $\pm$ 0.09 \\
25\%--30\% & 149.7 $\pm$ 6.0  & 353.0 $\pm$ 15.9 & 191.3 $\pm$ 14.9 & 2.56 $\pm$ 0.20 & 1.08 $\pm$ 0.10 \\
30\%--35\% & 124.8 $\pm$ 5.5  & 284.9 $\pm$ 13.2 & 153.9 $\pm$ 13.3 & 2.47 $\pm$ 0.22 & 1.08 $\pm$ 0.11 \\
35\%--40\% & 102.9 $\pm$ 5.1  & 227.1 $\pm$ 11.0 & 121.8 $\pm$ 11.7 & 2.37 $\pm$ 0.23 & 1.07 $\pm$ 0.12 \\
40\%--45\% & 83.2  $\pm$ 4.7  & 177.1 $\pm$ 8.8  & 96.0  $\pm$ 10.8 & 2.31 $\pm$ 0.27 & 1.08 $\pm$ 0.13 \\
45\%--50\% & 66.3  $\pm$ 4.3  & 136.5 $\pm$ 7.1  & 73.3  $\pm$ 8.9  & 2.21 $\pm$ 0.28 & 1.07 $\pm$ 0.14 \\
50\%--55\% & 52.1  $\pm$ 4.0  & 103.3 $\pm$ 6.5  & 55.5  $\pm$ 7.8  & 2.13 $\pm$ 0.32 & 1.07 $\pm$ 0.16 \\
55\%--60\% & 40.1  $\pm$ 3.8  & 76.0  $\pm$ 7.3  & 41.0  $\pm$ 6.6  & 2.04 $\pm$ 0.35 & 1.08 $\pm$ 0.20 \\
60\%--65\% & 30.1  $\pm$ 3.6  & 54.5  $\pm$ 7.1  & 30.2  $\pm$ 5.5  & 2.01 $\pm$ 0.40 & 1.11 $\pm$ 0.25 \\
\end{tabular} \end{ruledtabular}
\end{table*}

\begin{table*}[hbt]
\caption{\label{tab:auau062detdeta}
Transverse energy production results for 62.4 GeV Au$+$Au collisions. 
Listed are the total uncertainties (Type A, Type B, and Type C) for each 
centrality bin.}
\begin{ruledtabular} \begin{tabular}{rccccc}
Centrality & $<N_{\rm part}>$ & $<N_{qp}>$ & $\frac{dE_{T}}{d\eta}$ [GeV] & $\frac{1}{0.5<N_{\rm part}>}$ $\frac{dE_{T}}{d\eta}$ [GeV] & $\frac{1}{0.5<N_{qp}>}$ $\frac{dE_{T}}{d\eta}$ [GeV] \\\hline
0\%--5\%   & 342.6 $\pm$ 4.9 & 900.9 $\pm$ 21.7 & 389.7 $\pm$ 25.9 & 2.27 $\pm$ 0.13 & 0.87 $\pm$ 0.06 \\
5\%--10\%  & 291.3 $\pm$ 7.3 & 748.0 $\pm$ 20.4 & 320.5 $\pm$ 21.9 & 2.20 $\pm$ 0.13 & 0.86 $\pm$ 0.06 \\
10\%--15\% & 244.5 $\pm$ 8.9 & 614.7 $\pm$ 17.9 & 260.6 $\pm$ 18.8 & 2.13 $\pm$ 0.13 & 0.85 $\pm$ 0.07 \\
15\%--20\% & 205.0 $\pm$ 9.6 & 505.8 $\pm$ 16.9 & 212.1 $\pm$ 15.9 & 2.07 $\pm$ 0.13 & 0.84 $\pm$ 0.07 \\
20\%--25\% & 171.3 $\pm$ 8.9 & 414.3 $\pm$ 15.2 & 171.9 $\pm$ 14.4 & 2.01 $\pm$ 0.15 & 0.83 $\pm$ 0.08 \\
25\%--30\% & 142.2 $\pm$ 8.5 & 337.2 $\pm$ 12.5 & 138.6 $\pm$ 12.9 & 1.95 $\pm$ 0.16 & 0.82 $\pm$ 0.08 \\
30\%--35\% & 116.7 $\pm$ 8.9 & 271.1 $\pm$ 12.8 & 110.4 $\pm$ 11.7 & 1.89 $\pm$ 0.18 & 0.81 $\pm$ 0.09 \\
35\%--40\% & 95.2  $\pm$ 7.7 & 216.3 $\pm$ 11.0 & 86.9  $\pm$ 10.2 & 1.83 $\pm$ 0.19 & 0.80 $\pm$ 0.10 \\
40\%--45\% & 76.1  $\pm$ 7.7 & 168.8 $\pm$ 11.3 & 67.3  $\pm$ 8.7  & 1.77 $\pm$ 0.21 & 0.80 $\pm$ 0.12 \\
45\%--50\% & 59.9  $\pm$ 6.9 & 129.8 $\pm$ 9.7  & 51.2  $\pm$ 7.5  & 1.71 $\pm$ 0.23 & 0.79 $\pm$ 0.13 \\
50\%--55\% & 46.8  $\pm$ 5.2 & 98.8  $\pm$ 6.1  & 38.4  $\pm$ 6.4  & 1.64 $\pm$ 0.25 & 0.78 $\pm$ 0.14 \\
\end{tabular} \end{ruledtabular}
\end{table*}

\begin{table*}[htb]
\caption{
The slope parameters from a linear fit of $d\Et/d\eta$ as a function of 
$N_{qp}$, $d\Et/d\eta = a \times N_{qp} + b$ for each collision energy in 
Au$+$Au collisions. The value of $\chi^2$ has been calculated including 
Type A, Type B, and Type C uncertainties for each point.
}
\label{tab:detVsNq}
\begin{ruledtabular} \begin{tabular}{ccccccc}
& \sqsn (GeV) & $a$ (GeV) & $b$ (GeV) & $\chi^{2}$ & $n_{dof}$ & \\ \hline
& 200 & $0.617 \pm 0.023$ & $1.2 \pm 7.0$ & 0.098 & 9  & \\
& 130 & $0.551 \pm 0.020$ & $-2.1 \pm 6.5$ & 0.086 & 9  & \\
& 62.4 & $0.432 \pm 0.019$ & $-5.4 \pm 5.4$ & 0.163 & 9 & \\
\end{tabular} \end{ruledtabular}
\end{table*}

\begin{table*}[hbt]
\caption{\label{tab:auau200NcNpNqRatio}
Test of whether the ansatz, $[(1-x)\,\mean{\Npart}/2 +x\,\mean{\Ncoll} ]$, 
from Eq.~\ref{eq:crazy}, with $x=0.08$, is a proxy for \Nqp. The errors 
quoted on $\mean{N_{\rm part}}$, $\mean{N_{qp}}$, $\mean{\Ncoll}$ are 
correlated Type C and largely cancel in the $\mean{\Nqp}$/ansatz ratio.}
\begin{ruledtabular}
\begin{tabular}{rccccc}
Centrality & $\mean{N_{\rm part}}$ & $\mean{N_{qp}}$ & $\mean{\Ncoll}$ & ansatz & $\mean{\Nqp}$/ansatz \\
\hline
0\%--5\% &$ 350.9 \pm 4.7 $&$ 956.6 \pm 16.2 $&$ 1064.1 \pm 110.0 $& 246.5 & 3.88\\
5\%--10\% &$ 297.0 \pm 6.6 $&$ 789.8 \pm 15.3 $&$ 838.0 \pm 87.2 $& 203.7 & 3.88\\
10\%--15\% &$ 251.0 \pm 7.3 $&$ 654.2 \pm 14.5 $&$ 661.1 \pm 68.5 $& 168.3 & 3.89\\
15\%--20\% &$ 211.0 \pm 7.3 $&$ 540.2 \pm 12.3 $&$ 519.1 \pm 53.7 $& 138.6 & 3.90\\
20\%--25\% &$ 176.3 \pm 7.0 $&$ 443.3 \pm 10.4 $&$ 402.6 \pm 39.5 $& 113.3 & 3.91\\
25\%--30\% &$ 146.8 \pm 7.1 $&$ 362.8 \pm 12.2 $&$ 311.9 \pm 31.8 $& 92.5 & 3.92\\
30\%--35\% &$ 120.9 \pm 7.0 $&$ 293.3 \pm 11.0 $&$ 237.8 \pm 24.2 $& 74.6 & 3.93\\
35\%--40\% &$ 98.3 \pm 6.8 $&$ 233.5 \pm 9.2 $&$ 177.3 \pm 18.3 $& 59.4 & 3.93\\
40\%--45\% &$ 78.7 \pm 6.1 $&$ 182.7 \pm 6.8 $&$ 129.6 \pm 12.6 $& 46.6 & 3.92\\
45\%--50\% &$ 61.9 \pm 5.2 $&$ 140.5 \pm 5.3 $&$ 92.7 \pm 9.0 $& 35.9 & 3.91\\
50\%--55\% &$ 47.6 \pm 4.9 $&$ 105.7 \pm 5.5 $&$ 64.4 \pm 8.1 $& 27.0 & 3.91\\
55\%--60\% &$ 35.6 \pm 5.1 $&$ 77.3 \pm 6.8 $&$ 43.7 \pm 7.6 $& 19.9 & 3.89\\
60\%--65\% &$ 26.1 \pm 4.7 $&$ 55.5 \pm 7.1 $&$ 29.0 \pm 6.5 $& 14.3 & 3.87\\
65\%--70\% &$ 18.7 \pm 4.0 $&$ 39.0 \pm 6.7 $&$ 18.8 \pm 5.3 $& 10.1 & 3.86\\
70\%--75\% &$ 13.1 \pm 3.2 $&$ 27.0 \pm 4.9 $&$ 12.0 \pm 3.6 $& 7.0 & 3.86\\
75\%--80\% &$ 9.4 \pm 2.1 $&$ 19.0 \pm 3.2 $&$ 7.9 \pm 2.2 $& 5.0 & 3.83\\
80\%--92\% &$ 5.4 \pm 1.2 $&$ 10.3 \pm 1.5 $&$ 4.0 \pm 1.0 $& 2.8 & 3.67\\
$p$$+$$p$ & 2 &             $2.99\pm0.05       $&      1           &     1  &  2.99\\
\end{tabular}
\end{ruledtabular}
\end{table*}

\section{Extreme-Independent analyses in general}
\label{sec:extreme}
      
In Extreme Independent models for an $A$+$B$ nucleus-nucleus reaction, the 
nuclear geometry, i.e. the relative probability of the assumed fundamental 
elements of particle production, such as number of binary nucleon-nucleon 
(N+N) collisions (\Ncoll), nucleon participants or wounded nucleons 
(\Npart,WN), constituent-quark participants (NQP), or color-strings 
(wounded projectile quarks - AQM), can be computed from the assumptions of 
the model in a standard Glauber Monte Carlo 
calculation~\cite{Miller:2007ri} without reference to either the 
detector~\cite{MJTPRC69} or the particle production by the fundamental 
elements. Once the nuclear geometry is specified in this manner, it can be 
applied to the measured $p$$+$$p$ distribution (assumed equivalent to N+N) 
to derive the distribution (in the actual detector) of \Et or multiplicity 
(or other additive quantity) for the fundamental elementary collision 
process, i.e. a collision, a wounded nucleon (nucleon participant), 
constituent-quark participant or a wounded projectile quark 
(color-string), which is then used as the basis of the analysis of an 
$A$+$B$ reaction as the result of multiple independent elementary 
collision processes. The key experimental issue then becomes the linearity 
of the detector response to multiple collisions (better than 1\% in the 
present case), and the stability of the response for the different $A$+$B$ 
combinations and run periods used in the analysis. The acceptance of the 
detector is taken into account by making a correction for the probability, 
$p_0$, of measuring zero \Et for an N+N inelastic collision, which can 
usually be determined from the data~\cite{MJTPRC69} (as shown below).

The method for the calculation of the \Et distribution from an $A$+$B$ 
reaction in a given detector is illustrated for the \Ncoll or number of 
binary N+N collision model. The \Et distribution is equal to the sum:
\begin{equation}
\bigg({d\sigma\over d\Et}\bigg)_{\rm \Ncoll} = \sigma_{BA} \sum^{\rm N_{\rm max}}_{n=1} 
w_n\, P_n(\Et) 
\label{eq:wpnm}
\end{equation}
where $\sigma_{BA}$ is the measured $A$+$B$ cross section in the detector, 
$w_n$ is the relative probability for $n$ N+N collisions in the $A$+$B$ 
reaction with maximum value $n=\rm N_{\rm max}$, and $P_n(\Et)$ is the 
calculated \Et distribution on the detector for $n$ independent N+N 
collisions. If $f_1(\Et)$ is the measured \Et spectrum on the detector for 
an N+N collision that gives a nonzero \Et, and $p_0$ is the probability 
for an N+N collision to produce no signal in the detector (zero \Et) , 
then the correctly normalized \Et distribution for one N+N collision is:
\begin{equation}
P_1(\Et)= (1-p_0)f_1(\Et) +p_0 \delta(\Et),
\label{eq:P1}
\end{equation}
where $\delta(\Et)$ is the Dirac delta function and $\int f_1(\Et)\, 
d\Et=1$.  $P_n(\Et)$ (including the $p_0$ effect) is obtained by 
convoluting $P_1(\Et)$ with itself $n-1$ times
\begin{equation}
P_n(\Et) = \sum ^n_{i=0} {{n!} \over {(n-i)!\ i!} } \, 
p_0^{n-i} (1-p_0)^i f_i(\Et) 
\label{eq:wpnm2}
\end{equation}
where $f_0(\Et)\equiv\delta(\Et)$   
and $f_i(\Et)$ is the $i$-th convolution of $f_1(\Et)$:
 \begin{equation}
f_i (x)=\int_0^x dy\, f_1(y)\,f_{i-1}(x-y)\;\;\; . \label{eq:defcon} 
\end{equation}
Substituting Eq.~\ref{eq:wpnm2} into Eq.~\ref{eq:wpnm} and 
reversing the indices gives a form that is less physically 
transparent, but considerably easier to compute: 
\begin{equation}
\bigg({d\sigma\over d\Et}\bigg)_{\rm \Ncoll} = \sigma_{BA} \sum^{\rm N_{\rm max}}_{i=1} 
{w'}_i(p_0)\, f_i(\Et) 
\label{eq:wpnm3}
\end{equation}
where 
\begin{equation}
{w'}_i(p_0) = (1-p_0)^i\, \sum ^{\rm N_{\rm max}}_{n=i} {{n!} \over {(n-i)!\ i!} } \,
p_0^{n-i}\, w_n,  
\label{eq:wpnm4}
\end{equation}
which represents the weight (or relative probability) for $i$ convolutions 
of the measured $f_1(\Et)$ to contribute to the \Et spectrum in an $A$+$B$ 
collision, and where the term with ${w'}_{i=0}(p_0)$ in Eq.~\ref{eq:wpnm3} 
is left out because it represents the case when no signal is observed in 
the detector for an $A$+$B$ collision, i.e. ${w'}_{i=0}(p_0)=p_0^{BA}$. 
Note that the above example works for any other basic element of particle 
production e.g. constituent-quark-participant, if the labels NQP are 
substituted above for ``\Ncoll'' and ``N+N collision''. The method of 
determining $p_{0_{\rm NQP}}$ and $f_1^{\rm NQP}(\Et)$ will be described 
below.

In general the convolutions of $f_1(\Et)$ are performed analytically by 
fitting $f_1(\Et)$ to a Gamma distribution
\begin{equation}
f_1(x)=\frac{b}{\Gamma(p)} (bx)^{p-1} e^{-bx},
\label{eq:gammadist}
\end{equation}

where \[ p>0,\quad b>0,\quad 0\leq x\leq \infty \quad ,\] $\Gamma(p)$ is
is the Gamma function, which equals $(p-1)!$ if $p$ is an integer, and 
$\int_0^\infty f_1(x)\, dx=1$. The first few moments of the distribution 
are: \[\mu\equiv\mean{x}=\frac{p}{b} \quad \sigma=\frac{\sqrt{p}}{b} \quad
\quad \frac{\sigma^2}{\mu^2}=\frac{1}{p}\].
There are two reasons for this. In general the shape of \Et distributions 
in $p$$+$$p$ collisions is well represented by the Gamma distribution and 
the $n$-fold convolution (Eq.~\ref{eq:defcon}) is analytical
\begin{equation}
f_n(x)=\frac{b}{\Gamma(np)} (bx)^{np-1} e^{-bx},
\label{eq:gammaconv}
\end{equation}
i.e. $p\rightarrow np$ while $b$ remains unchanged. Notice that the mean 
$\mu_n$ and standard deviation $\sigma_n$ of the $n$-fold convolution obey 
the familiar rule:
\begin{equation}
\mu_n=n\mu, \quad \sigma_n=\sigma \sqrt{n}. 
\label{eq:convmoments}
\end{equation}

\subsection{The importance of collisions which give zero measured \Et}

The importance of taking account of $p_0$, the probability to give zero 
signal on the detector for an inelastic N+N collision (or other basic 
element of the calculation) can not be overemphasized. The properly 
normalized \Et distribution on the detector for one N+N collision is given 
by Eq.~\ref{eq:P1}, and the detected signal for $n$ independent N+N 
collisions is given by the binomial distribution, Eq.~\ref{eq:wpnm2}. The 
true detected mean for $n$ independent N+N collisions is $n$ times the 
true mean for one N+N collision, or:
\begin{equation} 
\langle \Et\rangle^{\rm true}_n=\int \Et\, P_n(\Et)\, d\Et=n \langle \Et\rangle^{\rm true},
\label{eq:meanETn}
\end{equation}
where  
\begin{eqnarray} 
\langle \Et\rangle^{\rm true}
&=&\int \Et\, P_1(\Et)\, d\Et \\ \nonumber
&=&(1-p_0) \int \Et\, f_1(\Et)\, d\Et \\ \nonumber
&=& (1-p_0) \mean{\Et}^{\rm ref},
\label{eq:defETref}
\end{eqnarray}
and $\mean{\Et}^{\rm ref}$ is the mean of the reference distribution, 
$f_1(\Et)$, the measured \Et spectrum for an N+N collision that gives 
nonzero \Et on the detector (Eq.~\ref{eq:P1}).  It is important to 
contrast Eq.~\ref{eq:meanETn} with the mean of the $n$-th convolution of 
the observed reference distribution, Eq.~\ref{eq:defcon},
\begin{eqnarray}
\langle \Et\rangle^{\rm ref}_n
&=&\int \Et\, f_n(\Et)\, d\Et \\ \nonumber
&=&n \langle \Et\rangle^{\rm ref}, 
\label{eq:meanETiobs}
\end{eqnarray}
which is $n$ times the observed reference $\mean{\Et}^{\rm ref}$, as it 
should be, and which differs from the mean of the true detected 
distribution, $P_n(\Et)$, for $n$ independently interacting projectile 
nucleons (Eq.~\ref{eq:meanETn}) by a factor of $1-p_0$ for all $n$, i.e.
\begin{eqnarray}
 \langle \Et\rangle^{\rm true}_n
&=& n \langle \Et\rangle^{\rm true} \\ \nonumber
&=&n (1-p_0) \mean{\Et}^{\rm ref}= (1-p_0) \langle \Et\rangle^{\rm ref}_n.
 \label{eq:true-refalln}
 \end{eqnarray}

\section{Application to the present data}
\label{sec:application}

As discussed in section \ref{sec:detector} above, the present measurements 
at \sqsn=200 GeV include Au$+$Au \Et distributions from the 2004 running 
period at RHIC and $p$$+$$p$ and $d$$+$Au distributions from the 2003 run.  
Although later runs with higher luminosity were tried, they suffer from 
tails due to pile-up of multiple interactions on the same event, which can 
be removed with fast electronics,\footnote{For continuous beams, in which 
fast triggered electronics are used with a short gate width, pile-up can 
be eliminated by a requirement that no additional interaction take place 
before or after the interaction of interest in a time interval 
corresponding to plus or minus the gate width~\cite{corplb}. Of course 
this requirement reduces the useful luminosity.} but which was not 
feasible with the present EMCal electronics~\cite{EMCalNIM}. This is most 
apparent for the $p$$+$$p$ data which is used as the measured \Et 
distribution, $f_1(\Et)$, for a {single} N+N collision.  The measured \Et 
distributions, with the requirement of a count (BBC$\geq 1$) in both the 
North and South BBC counters, are given as histograms of the number of 
counts in a given raw \Etemc bin such that the total number of counts sums 
up to the number of BBC counts (14,595,815 for $p$$+$$p$; 132,884,715 for 
Au$+$Au; 50,069,374 for $d$$+$Au). The distributions are then normalized 
to integrate to unity (Fig.~\ref{fig:f1}). Thus the normalized 
distributions are ``per BBC trigger per GeV'', so that the cross section 
$d\sigma/d\Et$ would be obtained for all distributions by multiplying by 
the relevant BBC cross section. This is not important for the $d$$+$Au or 
Au$+$Au data where the normalization is kept as the measured yield per BBC 
count per GeV in Au$+$Au or $d$$+$Au collisions, but is crucial for the 
$p$$+$$p$ measurement.  As discussed previously and tabulated in Table 
\ref{tab:corrections}, the correction scale factors are 6.68 for $p$$+$$p$, 
6.51 for $d$$+$Au and 6.87 for Au$+$Au, with Type C systematic 
uncertainties of $\sim\pm 6\%$ which are not relevant for the purposes of 
this analysis, except as an overall \Et scale uncertainty common to all 3 
distributions to which the absolute scale uncertainty of $\pm 1\%$ must be 
added.  We emphasize that these uncertainties are also common to 
all the calculations of the $d$$+$Au and Au$+$Au distributions to be 
presented, because they are based on the measured $p$$+$$p$ distribution. 
Note also that the detailed shape of \Et distributions has a 
slight dependence on the fiducial aperture due to statistical and 
dynamical fluctuations which are not taken into account by the simple 
scale correction. Thus an actual measurement in the reference acceptance 
will have slightly different upper tails in the region above the ``knee'' 
in the Au$+$Au distribution measured in the fiducial aperture 
$\Delta\eta\approx0.7, \Delta\phi\approx0.6\pi$ 
(Fig.~\ref{fig:f1}a)~\cite{egseeMJTRPP69,MJTRPP69}. Again this is not 
relevant to the present analysis in which the fiducial aperture is nearly 
identical for all three systems. \subsection{Determination of $p_0$ in the 
EMCal for an N+N collision} The requirement of the BBC$\geq 1$ trigger 
complicates the determination of the probability, $p_0$, of getting zero 
energy in the detector, in this case the EMCal, for an inelastic N+N 
collision, because it introduces a bias. For example, the high point 
clearly visible in the lowest bin of the $p$$+$$p$ data (Fig.~\ref{fig:f1}b) 
represents the events with zero \Et in the EMCal for a BBC trigger (in 
addition to the events with nonzero \Et in the lowest bin). This is a 
necessary quantity to measure but is not the same as $p_0$, the 
probability of getting zero \Et in the EMCal for an inelastic N+N 
collision. However, the BBC bias can be measured and corrected so that the 
cross section for \Et production in the EMCal in $p$$+$$p$ collisions can be 
determined; where we assume that $p$$+$$p$ and N+N are equivalent for \Et. 
This is the standard method used for all PHENIX $p$$+$$p$ cross section 
measurements in the EMCal, e.g. $\pi^0$~\cite{Adler:2003pb} and 
direct-$\gamma$~\cite{Adler:2006yt}, with details of the technique 
described in these references. The ratio of the measured \Et cross section 
per $p$$+$$p$ collision in the EMCal to the known $p$$+$$p$ inelastic cross 
section, then gives $1-p_0$~\cite{E802PRC63}.

The $p$$+$$p$ data are first fit to a Gamma distribution while expanding the 
error on the lowest data point by a factor of 10 so that it does not 
contribute to the fit. The Gamma distribution integrates to a fraction 
$Y_{\Gamma}^{pp}$ of the number of BBC triggers. Then the observed yield 
per BBC count is converted to the observed cross section by multiplying by 
the measured BBC cross section of $\sigma_{\rm BBC}=$ 23.0 mb $\pm 9.7\%$. 
This cross section must then be corrected for the BBC bias, 
$1-\varepsilon_{\rm bias}$, the probability of getting no BBC count when 
there is finite energy in the central spectrometer. This was measured 
using clock triggers for single charged particles in the central 
spectrometer as well as from the ratio of the yield of high $p_T$ $\pi^0$ 
with and without the BBC$\geq 1$ trigger~\cite{Adler:2006yt} and found to 
be a constant $\varepsilon_{\rm bias}=0.79\pm 0.02$, independent of $p_T$.  
Thus, the measured \Et cross section per $p$$+$$p$ collision equals 
$Y_{\Gamma}^{pp} \times \sigma_{\rm BBC}/\varepsilon_{\rm bias}$.
The probability of detecting zero \Et in the detector for an inelastic N+N 
collision is then computed from the ratio of the integrated cross section 
of the measured \Et distribution to the 42 mb $p$$+$$p$ inelastic cross 
section, $\sigma_{\rm inel}$:

\begin{eqnarray}
\label{eq:1-p0}
1-p_0
&=&{1 \over \sigma_{\rm inel}}\, {23.0\,{\rm mb}\pm 9.7\%  
\over 0.79\pm 0.02} \, Y_{\Gamma}^{pp}  \\ \nonumber
&=&0.693 (\pm 10\%) \, Y_{\Gamma}^{pp}.
\end{eqnarray}  	

\begin{figure}[htb] 
\includegraphics[width=1.0\linewidth]{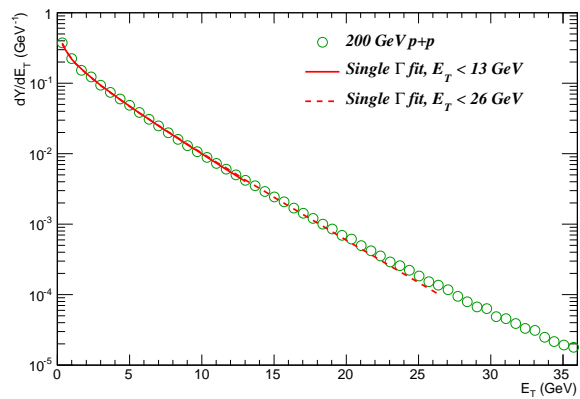} 
\caption{(Color online) 
Fits of the $p$$+$$p$ data to a single $\Gamma$ distribution for the ranges 
$\Et<13.3$ and $\Et<26.6$ GeV.}
      \label{fig:SingleGammapp}
   \end{figure}

The procedure is a two-step process. First the fit is performed with the
error in the lowest bin increased by a factor of 10, so that the counts
with zero \Et do not distort the fit. Then trial values of
$Y_{\Gamma}^{pp}$ and $1-p_0$ are derived from Eq.~\ref{eq:1-p0} and the
data are corrected to a data set for which the lowest bin in the
distribution is replaced by the fitted value in this bin and the original
error is restored, so that the distribution, $d Y/d\Et$ which previously
integrated to unity, now integrates to $Y_{\Gamma}^{pp}$. This data set is
then refit for the final results. The value of $1-p_0$ is evaluated from
the new $Y_{\Gamma}^{pp}$ which, with the procedure indicated, typically
does not differ significantly from the trial value. The parameters for the
fit of the $p$$+$$p$ data to a Gamma distribution are given in
Table~\ref{table:Gammafitpp}. Only the data for $\Et<13.3$ GeV are used in
the fit and the following analysis to avoid influence from the
tail, which is presumed to be from residual pile-up. However, the fit was
also extended to $\Et<26.6$ GeV as a systematic check
(Fig.~\ref{fig:SingleGammapp}).
       
The poor $\chi^2_{\rm min}$/dof for both fits has at least two sources. 
For low \Et, the statistical uncertainties with millions of events per bin 
are $\sim 1/1000$ so any uncorrected few percent systematic effect for 
each data point (e.g. such as not bin-shifting for the falling spectrum) 
gives a large contribution to the $\chi^2$. At larger $\Et>20$ GeV, the 
data clearly lie above the fit, which is emphasized by the fit with 
$\Et<26.6$ GeV. This difference is presumed to be due to residual pile-up. 
In any case, the fits for both \Et ranges follow the $p$$+$$p$ data for more 
than two orders of magnitude and have $\mean{\Et^{\rm ref}}$ which differ 
by 0.6\%, so are more than adequate for the multiple collision 
calculations, for which the dominant effect in convolutions is the mean 
value. An 0.6\% variation in $\mean{\Et^{\rm ref}}$ will result in an 
0.6\% change in the \Et scale of the calculations which is negligible 
compared to the dominant systematic uncertainty to be discussed below.  
The tail only enters when the geometry is exhausted~\cite{E802PRC63}, 
which is not reached for the present $d$$+$Au and Au$+$Au data. Following 
the standard practice, the uncertainties on the fitted parameters, 
$Y_{\Gamma}^{pp}$, $b$ and $p$, in Table~\ref{table:Gammafitpp} have been 
increased by a factor of $\sqrt{4866/17}=16.9$ and $\sqrt{6715/37}=13.4$, 
respectively. Thus, the fractional statistical uncertainty on $1-p_0$ from 
the fitted $Y_{\Gamma}^{pp}$ is $0.006/0.933= 0.6\%$, which is still small 
compared to the uncertainties on the parameters in Eq.~\ref{eq:1-p0} of 
which the 9.7\% uncertainty in the BBC cross section is predominant. 
Adding the 0.6\% fractional uncertainty in quadrature with the two 
fractional uncertainties on the parameters in Eq.~\ref{eq:1-p0} gives a 
total systematic uncertainty on $1-p_0$ of 10.1\%. Thus, the values of 
$1-p_0$ are taken as 0.647, 0.660, with a systematic uncertainty of 10\% 
as indicated in Table~\ref{table:Gammafitpp}.

\subsection{Calculations of the various models}

\begin{figure}[htb] 
\includegraphics[width=1.0\linewidth]{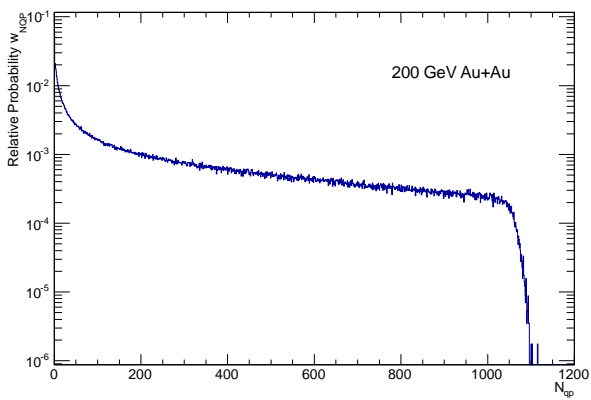} 
\caption{(Color online) 
Distribution of the Number of Quark Participants in Au$+$Au at 
$\sqrt{s_{\rm NN}}=200$ GeV}
\label{fig:NPQdist}
\end{figure}

The starting point requires the relative probabilities, $w_n$, for the 
number of binary N+N collisions, nucleon participants, constituent-quark 
participants from q-q scattering (NQP), and wounded projectile quarks from 
q-N scattering (AQM) for $\sqrt{s_{\rm NN}}=200$ GeV $p$$+$$p$, $d$$+$Au, 
Au$+$Au collisions. These were calculated by the standard Glauber Monte 
Carlo method, as described in section~\ref{sec:Glauber}. For Au$+$Au they 
are plotted in Fig.~\ref{fig:NPQdist}.

\label{sec:KR}    

There was no explicit AQM calculation in Au$+$Au; the probability for $n$ 
wounded projectile quarks was taken to be the sum of the probabilities for 
$2n$ and $2n-1$ constituent-quark participants. The weights for $p$$+$$p$ 
and $d$$+$Au are tabulated in Tables.~\ref{table:qqwts} and 
\ref{table:origwts-AuAu}. The weights in these tables are defined as the 
`original' weights, ($p_0=0$, $\epsilon\equiv 1-p_0=1.0$), before 
correction for $p_0$.

   \begin{table*}[htb]
\caption{
Fitted parameters $Y_{\Gamma}^{pp}$ $b$, $p$ of $p$$+$$p$ data, and 
calculated $1-p_0$. Note that the standard errors on these parameters 
using $\chi^2=\chi^2_{\rm min}+1$ have been multiplied by 
$\sqrt{\chi^2_{\rm min}/{\rm dof}}$ in each case.}
\label{table:Gammafitpp}
\begin{ruledtabular}
\begin{tabular}{ccccccc}
System &$Y_{\Gamma}^{pp}$ & $b$ (GeV)$^{-1}$ & $p$ & $\mean{\Et}^{\rm ref}$ GeV &$\chi^2_{\rm min}$/dof& $1-p_0$\\[0.2pc]\hline
$p$$+$$p$ $\Et<13.3$ & $0.933\pm 0.006$  & $0.273\pm 0.003$ & $0.724\pm  0.010$ & 2.64 &4866/17 & $0.647\pm 0.065$ \\
$p$$+$$p$ $\Et<26.6$ & $0.952\pm 0.004$  & $0.263\pm 0.003$ & $0.692\pm  0.007$ & 2.63 &6715/37 & $0.660\pm 0.066$ \\
\end{tabular}
\end{ruledtabular}

\caption{
Original weights $w_n$ ($p_0=0$, $\epsilon\equiv 1-p_0=1.0$) for $p$$+$$p$ 
and $d$$+$Au at $\sqrt{s}=200$ GeV. Note that $\sigma=9.36$ mb was used 
for q-q scattering to obtain a N+N $\sigma^{\rm inel}=42.0$ mb. 
These AQM weights come from the q-q scattering calculation tabulated from 
the distribution of projectile participants, NQP (p), NQP(deuteron). 
The symbol ``$...$'' in the table indicates additional weights for $n\geq 7$.
\label{table:qqwts}}
\begin{ruledtabular}
\begin {tabular}{ccccccc}
    & \multicolumn{2}{c}{$p$$+$$p$}    
& \multicolumn{2}{c}{$d$$+$Au}  
& \multicolumn{2}{c}{Au$+$Au} 
\\
  $n$ 
& NQP 
& AQM 
& NQP 
& AQM 
& \Npart 
& NQP 
\\ \hline
1 & 0.00   & 0.609  & 0.00   & 0.131  & 0.00   & 0.00   \\
2 & 0.465  & 0.285  & 0.0867 & 0.124  & 0.0660 & 0.0613 \\
3 & 0.238  & 0.106  & 0.0516 & 0.202  & 0.0304 & 0.0204 \\
4 & 0.169  &        & 0.0529 & 0.0925 & 0.0269 & 0.0209 \\
5 & 0.0946 &        & 0.0473 & 0.118  & 0.0220 & 0.0176 \\
6 & 0.0333 &        & 0.0451 & 0.332  & 0.0195 & 0.0157 \\
7 &        &        &  ...   &        &  ...   & ...    \\
\end{tabular}
\end{ruledtabular}


\caption{
Original \Ncoll and AQM weights $w_n$ ($p_0=0$, $\epsilon\equiv 
1-p_0=1.0$) for Au$+$Au at $\sqrt{s}=200$ GeV. Note that $\sigma=9.36$ mb 
was used for q-q scattering to obtain a N+N $\sigma^{\rm inel}=42.0$ mb. 
The symbol ``$...$'' in the table indicates additional weights for 
$n\geq{7}$.
\label{table:origwts-AuAu}}
\begin{ruledtabular}
\begin {tabular}{ccccc}
& $n$  & Au$+$Au \Ncoll &  Au$+$Au AQM  & \\ 
\hline
& 1    & 0.0660 & 0.0613 & \\
& 2    & 0.0405 & 0.0414 & \\
& 3    & 0.0287 & 0.0333 & \\
& 4    & 0.0232 & 0.0263 & \\
& 5    & 0.0191 & 0.0214 & \\
& 6    & 0.0169 & 0.0184 & \\
& 7    &  ...     & ...  & \\
\end{tabular}
\end{ruledtabular}
\end{table*}


\subsubsection{Correction of the weights $w_n$ to ${w'}_i(p_0)$ for 
\Npart, NQP and AQM in $p$$+$$p$ to account for $p_0$}
    
Because the $p_0$ is calculated for a $p$$+$$p$ collision, one has to 
recompute the $p$$+$$p$ weights in each model to find the $p_{0_{\rm AQM}}$, 
$p_{0_{\rm NQP}}$, and $p_{0_{\rm N_{\rm part}}}$ so that the new weights 
for the elementary processes sum up to $1-p_0$ for the $p$$+$$p$ 
collision. The value $1-p_0=0.647$ for $p$$+$$p$ collisions, from 
Table~\ref{table:Gammafitpp}, gives the probability $p_0=0.353$ for an 
inelastic N+N collision to give zero energy into our acceptance, i.e. zero 
detected \Etemc.  For \Ncoll, which is based on N+N collisions, $p_0$ is 
simply that of a $p$$+$$p$ collision. For \Npart, because a $p$$+$$p$ collision 
is 2 participants, it is assumed that both participants had equal 
$p_{0_{\rm N_{\rm part}}}$, and so the case when only 1 WN deposited 
energy is not counted. This is done because both BBCs are required to 
count on a N+N collision although there are certainly cases when both WN 
could give a BBC count but only 1 would give a nonzero \Et. If the case 
when only 1 WN deposited energy were allowed, then the only way to get 
zero energy on a $p$$+$$p$ collision is for both WN to give zero energy i.e. 
$p_{0_{\rm WN}}=p_0^2=0.125$, $\epsilon_{\rm WN}=0.875$, but then the 
weight for 1 WN would have to be included in this calculation. We chose 
instead to require both WN to deposit energy, hence a $p$$+$$p$ collision 
equaled 2 WN, i.e. $\epsilon_{pp}=1-p_0=\epsilon_{\rm WN}^2$, so 
$\epsilon_{\rm WN}=\sqrt{1-p_0}=0.804$.

For NQP, Eq.~\ref{eq:wpnm4} was used to calculate the value of 
${w'}_{i=0}(p_{0_{\rm NQP}})$ for any $p_{0_{\rm NQP}}$ with the case 
NQP=1 not allowed, so 
${w'}_{i=0}(p_{0_{\rm NQP}})+{w'}_{i=1}(p_{0_{\rm NQP}})=p_0=0.353$ 
was solved, with result $\epsilon_{\rm QP}=1-p_{0_{\rm NQP}}=$0.659. For 
the AQM, the total efficiency of the projectile quarks (color-strings) 
should add up to the efficiency of a $p$$+$$p$ collision at midrapidity. 
Thus the equation ${w'}_{i=0} (p_{0_{\rm AQM}})=p_{0}=1-0.647 =0.353$ was 
solved, with result $\epsilon_{\rm AQM}=1-p_{0_{\rm AQM}}=0.538$.
    
Note that there can be confusion in the AQM model because in a $p$$+$$p$ 
collision, represented as 1 to 3 q+$p$ collisions, the struck proton may 
have the efficiency of a Wounded Nucleon rather than that of a Wounded 
Projectile Quark. Such an asymmetric AQM model can be calculated. However, 
if one thinks of the AQM model as the number of color strings rather than 
number of wounded projectile quarks, then the detection efficiency, 
$\epsilon_{\rm AQM}=1-p_{0_{\rm AQM}}=0.538$, can be thought of as the 
detection efficiency for a color string.
   
\subsubsection{Correcting the $p$$+$$p$, $d$$+$Au and Au$+$Au weights.}

Applying $p_{0_{\rm AQM}}$, $p_{0_{\rm NQP}}$, and $p_{0_{\rm WN}}$ to 
correct the $p$$+$$p$, $d$$+$Au and Au$+$Au weights is straightforward 
and given by Eq.~\ref{eq:wpnm4}. The weights from Tables~\ref{table:qqwts} 
and \ref{table:origwts-AuAu} corrected for these efficiencies are 
summarized in Tables~\ref{table:corrwts} and \ref{table:corrwts-AuAu}.

\begin{table*}[htb]
\caption{
Corrected weights ${w'}_{i}$ for $p$$+$$p$, $d$$+$Au Au$+$Au,at $\sqrt{s}=200$
GeV. Note that $1-p_0$ is the sum of the weights in the column (including
weights not tabulated) and is the not the BBC efficiency, but the
probability to get a nonzero \Etemc on an $A$+$B$ collision.
\label{table:corrwts}}
\begin{ruledtabular}
\begin {tabular}{ccccccc}
    & \multicolumn{2}{c}{$p$$+$$p$}    
& \multicolumn{2}{c}{$d$$+$Au}  
& \multicolumn{2}{c}{Au$+$Au} 
\\
& NQP
& AQM
& NQP
& AQM
& \Npart
& AQM
\\
$n$
& $\epsilon_{\rm NQP}$=0.659 
& $\epsilon_{\rm AQM}$=0.538 
& $\epsilon_{\rm NQP}$=0.659 
& $\epsilon_{\rm AQM}$=0.538 
& $\epsilon_{\rm WN}$=0.804
& $\epsilon_{\rm NQP}$=0.659 
\\
\hline
1       & 0.00    & 0.506  & 0.00   & 0.259   & 0.00    & 0.00   \\
2       & 0.378   & 0.125  & 0.0918 & 0.251   & 0.0596  & 0.0474 \\
3       & 0.173   & 0.0164 & 0.0726 & 0.199   & 0.0333  & 0.0270 \\
4       & 0.0731  &        & 0.0664 & 0.120   & 0.0277  & 0.0231 \\
5       & 0.0202  &        & 0.0601 & 0.0467  & 0.0230  & 0.0195 \\
6       & 0.00272 &        & 0.0552 & 0.00802 & 0.0199  & 0.0168 \\
7       &         &        & ...    &         &  ...    &  ...   \\
\\
1-$p_0$ & 0.647   & 0.647  & 0.926  & 0.883   &   0.973 &  0.956 \\
\end{tabular}
\end{ruledtabular}


\caption{
Corrected weights ${w'}_{i}$ for Au$+$Au at $\sqrt{s}=200$ GeV. Note that 
$1-p_0$ is the sum of the weights in the column (including weights not 
tabulated) and is the not the BBC efficiency, but the probability to get a 
nonzero \Etemc on the $A$+$B$ collision.
\label{table:corrwts-AuAu}}
\begin{ruledtabular}
\begin {tabular}{ccccc}
&   & \Ncoll &  AQM  \\ 
&  $n$ &$\epsilon_{\rm coll}$=0.647&$\epsilon_{\rm AQM}$=0.538 & \\ 
\hline
& 1    & 0.0723 & 0.0756 & \\
& 2    & 0.0433 & 0.0494 & \\
& 3    & 0.0312 & 0.0362 & \\
& 4    & 0.0247 & 0.0284 & \\
& 5    & 0.0205 & 0.0235 & \\
& 6    & 0.0175 & 0.0202 & \\
& 7    &  ...     & ...      & \\
\\ 
& 1-$p_0$&   0.970  &  0.958   & \\
\end{tabular}
\end{ruledtabular}


\caption{
Parameters $b$, $p$ of the element indicated from the fit to $p$$+$$p$ data, 
cut for $\Et<13.3$ GeV ($\Etemc<2$ GeV). $Y_{\Gamma}^{fit}$ is the fitted 
integral of the $p$$+$$p$ distribution. For \Ncoll, the fit is a single 
$\Gamma$ to the $p$$+$$p$ distribution from which $\epsilon_{pp}$ is 
calculated; 
for \Npart, $p_{WN}=p_{pp}/2$, $\epsilon_{WN}=\sqrt{\epsilon_{pp}}$. For 
NQP and AQM the fits are the deconvolution of elements with weights 
${w'}_{i}$ which do not sum to unity but sum to $\epsilon_{pp}=0.647$ so 
that $Y_{\Gamma}^{pp}=Y_{\Gamma}^{fit}\times \epsilon^{pp}$=0.948 (NQP), 
0.944 (AQM), a good check (within 1.6\% and 1.1\% respectively).
\label{table:Gammafits}}
\begin{ruledtabular} \begin {tabular}{cccccccc}
Model & $\epsilon_{element}$ & element &$Y_{\Gamma}^{fit}$ & $b$ (GeV)$^{-1}$ & $p$ & $\langle \Et\rangle^{\rm ref}_{elem}$ (GeV)& $\langle \Et\rangle^{\rm true}_{elem}$ (GeV)\\
\hline
\Ncoll & 0.647  & $p$$+$$p$   & 0.933  & 1.83/6.68 & 0.723 & $2.64$ & 1.71\\
\Npart & 0.804  & 1 WN      & 0.933  & 1.83/6.68 & 0.363 & $1.32$ & 1.06\\
NQP   & 0.659  & 1 QP      & 1.466  & 2.00/6.68 & 0.297 & $0.994$ & 0.655\\
AQM   & 0.538  & 1 string  & 1.460  & 2.10/6.68 & 0.656 & $2.09$ & 1.12\\
\end{tabular} \end{ruledtabular}
\end{table*}

\subsubsection{Derivation of the \Et distribution of the basic elements 
from the $p$$+$$p$ \Et distributions followed by calculation of the $d$$+$Au 
and Au$+$Au distributions}

At this point the raw \Etemc distributions in the fiducial aperture had 
been corrected to the total hadronic $\Et=d\Et/d\eta|_{\eta=0}$ by making 
a change of scale from \Etemc to \Et by the correction factors of 6.68 for 
$p$$+$$p$, 6.51 for $d$$+$Au and 6.87 for Au$+$Au (Fig.~\ref{fig:f1}). The 
$p$$+$$p$ and the elementary WN, NQP, AQM distributions $f_i(\Et)$ in 
Eqs.~\ref{eq:P1}--\ref{eq:wpnm3} are taken as $\Gamma$ distributions and 
then the $p$$+$$p$ distribution (Fig.~\ref{fig:f1}b) is deconvoluted using 
the efficiency corrected weights, ${w'}_{i}$, to find the parameters of 
the elementary NQP, or AQM distributions.  For the WN the deconvolution 
from $p$$+$$p$ is analytical.

\begin{figure}[htb] 
      \centering
      \includegraphics[width=1.0\linewidth]{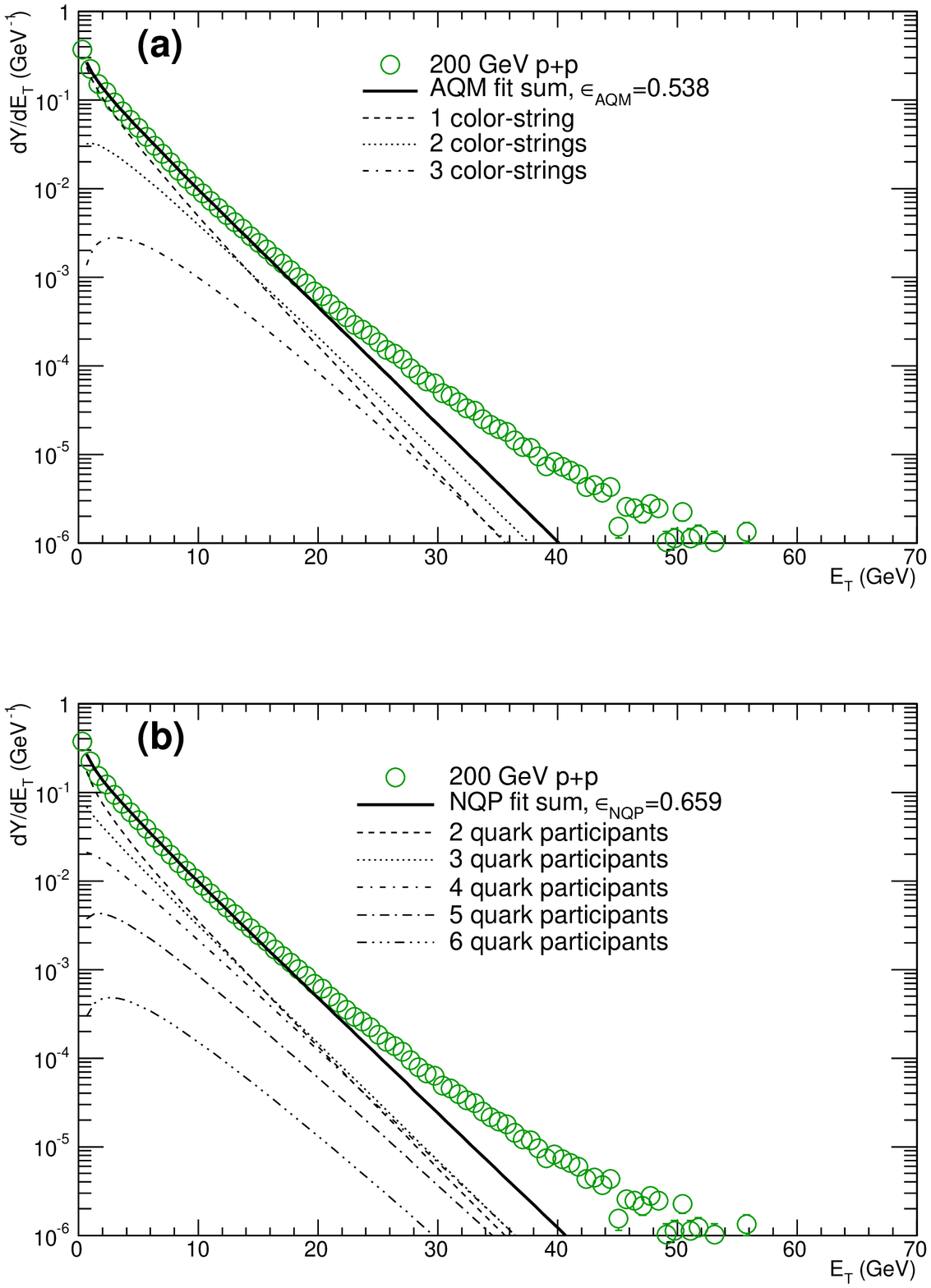} 
      \caption{(Color online) 
(a) Deconvolution fit to the $p$$+$$p$ $\Et$ distribution for $\Et<13.3$ GeV 
at $\sqrt{s_{\rm NN}}=200$ GeV with the corrected weights ${w'}^{\rm 
AQM}_{i}$ calculated in the Additive Quark model (AQM) using the symmetric 
color-string efficiency, $\epsilon_{\rm AQM}=1-p_{0_{\rm AQM}}=0.538$. 
Lines represent the properly weighted individual \Et distributions for 
1,~2,~3 color-strings plus the sum.  On the y-axis intercept, the top line is 
the sum and the lower curves in descending order are the \Et distributions 
of 1,2,3 color-strings. (b) Deconvolution fit to the same $p$$+$$p$ \Et 
distribution for $\Et<13.3$ GeV with the corrected weights 
${w'}^{\rm NQP}_{i}$ with $\epsilon_{NQP}=1-p_{0_{\rm NQP}}=0.659$ 
calculated in the NQP model. Lines represent the properly weighted 
individual \Et distributions for the underlying 2,~3,~4,~5,~6 
constituent-quark participants plus the sum. }

\label{fig:ppdecon}
   \end{figure}

The results of the fit to a single $\Gamma$ distribution ($p$$+$$p$) were 
given in Table ~\ref{table:Gammafitpp} and Fig.~\ref{fig:SingleGammapp}. 
The deconvolution of $p$$+$$p$ to sums of elementary $\Gamma$ distributions 
with AQM and NQP weights ${w'}_{i}$ are shown in Fig.~\ref{fig:ppdecon} 
and given in Table~\ref{table:Gammafits}.

\begin{figure}[htb] 
\includegraphics[width=1.0\linewidth]{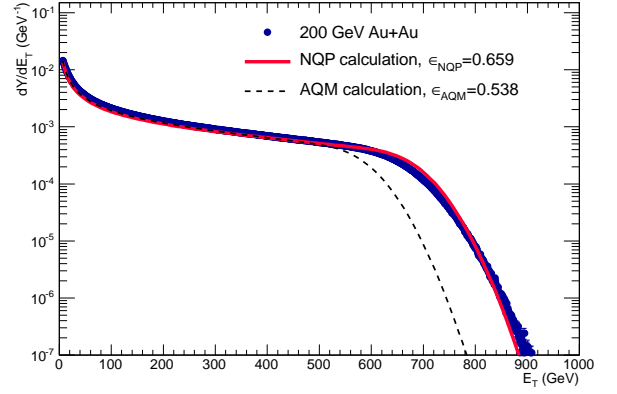} 
\caption{(Color online) 
\Et distributions at $\sqrt{s_{\rm NN}}=200$ GeV calculated in the Number 
of constituent-Quark Participants or NQP model, with 
$\epsilon_{NQP}=1-p_{0_{\rm NQP}}=0.659$ for Au$+$Au together with the AQM 
calculations with efficiencies indicated.}
\label{fig:all}
\end{figure}

These parameters are then used in Eq.~\ref{eq:wpnm3} with the $d$$+$Au and 
Au$+$Au corrected weights to compute the $\Et$ distributions for these 
systems. The results for the Additive Quark Model (AQM) using the above 
$\epsilon_{\rm AQM}=1-p_{0_{\rm AQM}}=0.538$ and the constituent-Quark 
Participant (NQP) model with $\epsilon_{NQP}=1-p_{0_{\rm NQP}}=0.659$ are 
shown for Au$+$Au in Fig.~\ref{fig:all}. Both the shape and magnitude of 
the calculation with the NQP model are in excellent agreement with the 
entire Au$+$Au measurement including the upper edge of the calculation, 
which is essentially on top of the measured \Et distribution, well within 
the principal $\pm 10\%$ systematic uncertainty in $1-p_0$ from the BBC 
cross section (Eq.~\ref{eq:1-p0}). This uncertainty is common to both AQM 
and NQP calculations so does not affect the difference in the AQM and NQP 
curves, both curves scale together in \Et by the same $\pm10.1$\% with 
respect to the data. Another advantage of the Extreme Independent Models 
is that all the calculations are based on the measured data. Thus the 6\% 
Type C common systematic uncertainty on the absolute \Et scale 
(Table~\ref{tab:sysErrors}) cancels in relative comparisons of the data to 
the calculations---all the curves and the data scale together by the same 
fraction in \Et.

Interestingly, the AQM model is not identical to the NQP model for the 
symmetric Au$+$Au system, but 12\% lower in the \Et knee. This is due to 
the $p_0$ effect in the $p$$+$$p$ collision, which has different effects on 
the AQM and NQP calculations. This was checked by repeating the AQM 
(color-string) and NQP calculations with $1-p_0=1.0$ detection efficiency 
in a $p$$+$$p$ collision to confirm that the AQM and NQP models really do 
give identical results in symmetric Au$+$Au collisions for 100\% 
efficiency.

\begin{figure}[htb] 
\includegraphics[width=1.0\linewidth]{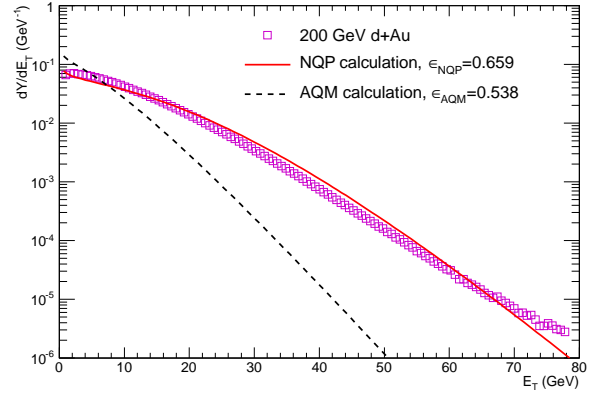} 
\caption{(Color online) 
$d$$+$Au measurements compared to the AQM and NQP model calculations.}
\label{fig:dAufits}
\end{figure}

\begin{figure}[htb] 
      \centering
      \includegraphics[width=1.0\linewidth]{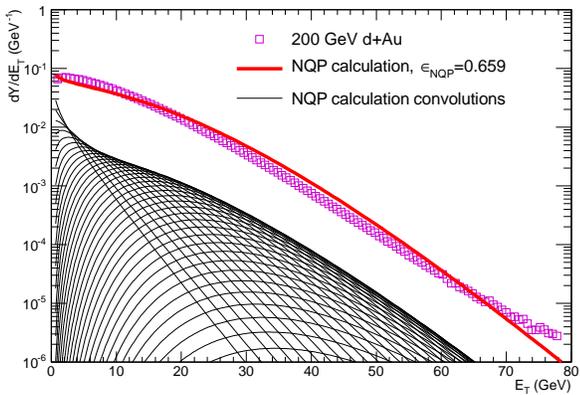} 
      \caption{(Color online) 
\Et distributions at $\sqrt{s_{\rm NN}}=200$ GeV in $d$$+$Au calculated in 
the Quark Participant (NQP) model with $\epsilon_{\rm NQP}=1-p_{0_{\rm 
NQP}}=0.659$ together with the individual visible convolutions for NQP, 
i.e. 2,3,\ldots 33, out of a maximum of 50 NQP considered. }
      \label{fig:f4}
\end{figure}

The major difference in the NQP and AQM calculations with respect to the 
measurements shows up in the asymmetric $d$$+$Au system, 
Fig.~\ref{fig:dAufits}, where the NQP calculation closely follows the 
$d$$+$Au \Et distribution in shape and in magnitude over a range of a 
factor of 1000 in cross section.  The AQM calculation disagrees both in 
shape and magnitude, with a factor of 1.7 less transverse energy emission 
than in the measurement.  This clearly indicates the need for emission 
from additional quark participants in the Au target beyond those in the 
deuteron, as shown by the individual components of the NQP calculation for 
$d$$+$Au (Fig.~\ref{fig:f4}).  It is also clear that having the comparison 
between the NQP and AQM models for asymmetric systems is crucial in 
distinguishing the models.

Previously, the hypothesis of quark-participant scaling in Au$+$Au 
collisions had been tested only for mean values by plotting 
$\mean{d\Et/d\eta}/(\mean{N_{qp}}/2)$ vs 
${\Npart}$~\cite{EreminVoloshinPRC67,DeBhattPRC71,NouicerEPJC49} as 
applied here in Fig.~\ref{fig:detNormNquark}. The present work extends the 
NQP model to {\em distributions}, as described in 
section~\ref{sec:application} and shown in Fig.~\ref{fig:all}. By doing 
so, we are able to make a crucial consistency check---the 
$\mean{d\Et/d\eta}/{N_{qp}}=0.617\pm0.023$ GeV from the linear fit 
(Fig.~\ref{fig:detdetaNquark}) in Au$+$Au is equal (within $<1$ standard 
deviation) to the value $\mean{\Et}^{\rm true}_{qp}=0.655\pm 0.066$ GeV 
derived for a quark-participant from the deconvolution of the $p$$+$$p$ \Et 
distribution (Table~\ref{table:Gammafits}).


         \begin{figure}[htb] 
      \includegraphics[width=1.0\linewidth]{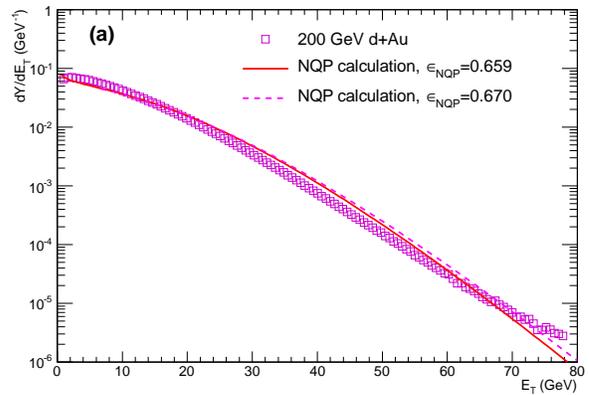}   
      \includegraphics[width=1.0\linewidth]{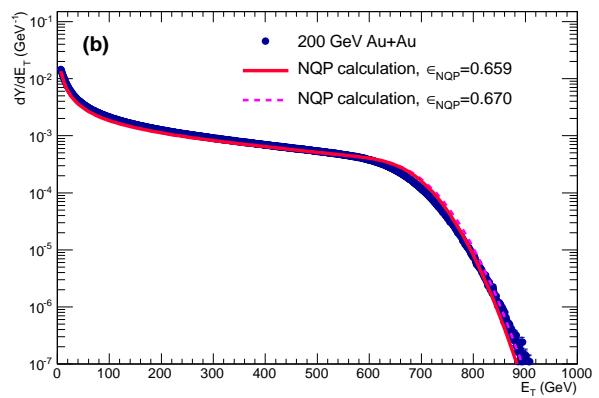} 
      \caption{(Color online)
Systematic checks of $\Et\equiv d\Et/d\eta|_{y=0}$ calculations using 
$p$$+$$p$ fits with $\Et<26.6$ GeV (a) $d$$+$Au data compared to standard 
calculation in the NQP model with $\epsilon_{\rm NQP}=1-p_{0_{\rm 
NQP}}=0.659$, for $1-p_0=0.647$ in a $p$$+$$p$ collision from fit with 
$\Et<13.3$ GeV compared to $\epsilon_{\rm NQP}=1-p_{0_{\rm NQP}}=0.670$ 
for $1-p_0=0.660$ when the fit to the $p$$+$$p$ data is extended to 
$\Et<26.6$ GeV. (b) Au$+$Au calculation for the same conditions as 
$d$$+$Au in (a).}
      \label{fig:syscheck}
\end{figure}

\begin{figure}[htb] 
\includegraphics[width=1.0\linewidth]{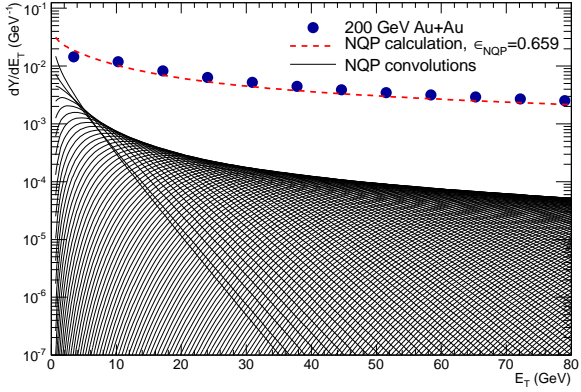} 
\caption{(Color online) 
Measured \Et distribution in Au$+$Au at $\sqrt{s_{\rm NN}}=200$ GeV on the 
same \Et scale as Fig.~\ref{fig:f4} compared to the calculation in the 
Quark Participant (NQP) model with $\epsilon_{\rm NQP}=1-p_{0_{\rm 
NQP}}=0.659$ together with the individual visible convolutions for NQP in 
this \Et range, i.e. 2,3, \ldots 114, out of 584 convolutions with visible 
contribution to the full distribution, out of a maximum of 1020 NQP 
considered.}
\label{fig:v12p5b}
  \end{figure}

         \begin{figure}[htb] 
      \includegraphics[width=1.0\linewidth]{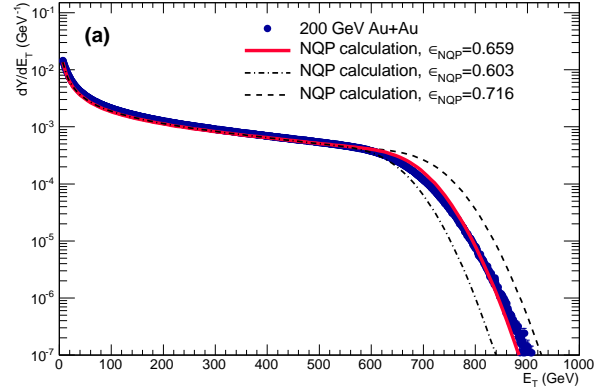}
      \includegraphics[width=1.0\linewidth]{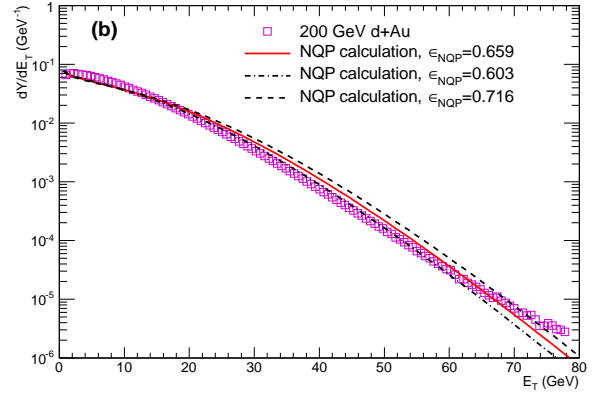} 
      \caption{(Color online)
$\Et\equiv d\Et/d\eta|_{y=0}$ distributions at $\sqsn=200$ GeV: (a) 
Au$+$Au compared to the NQP calculations using the central $1-p_0=0.647$ 
and $\pm 1\sigma$ variations of $1-p_0=0.582,0.712$ for the probability of 
getting zero \Et on a $p$$+$$p$ collision with resulting $\varepsilon_{\rm 
NQP}=0.659,0.603,0.716$, respectively.  (b) $d$$+$Au calculation for 
the same conditions as in (a).}
      \label{fig:dAuAufinal}
\end{figure}

\begin{figure}[htb] 
\includegraphics[width=1.0\linewidth]{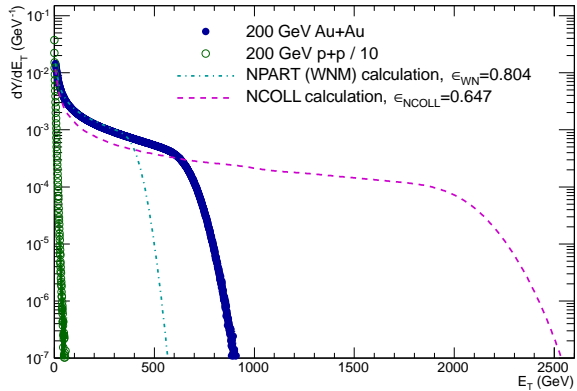} 
\caption{(Color online) 
Au$+$Au measurement of $d\Et/d\eta$ compared to the \Npart-WNM (dot-dash) 
and \Ncoll (dashes) model calculations.}
\label{fig:v12p4bot}
\end{figure}

\begin{figure}[htb] 
\includegraphics[width=1.0\linewidth]{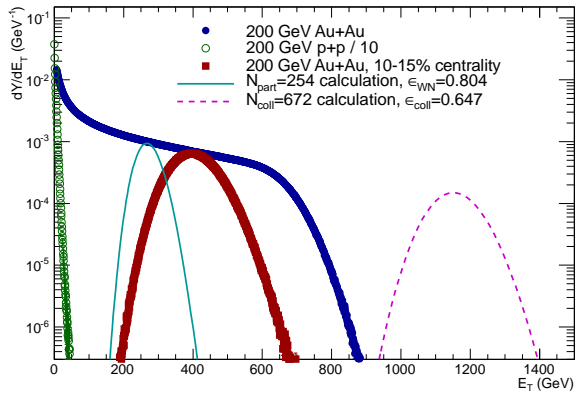} 
\caption{(Color online) 
Au$+$Au measurement of $d\Et/d\eta$, with 10\%--15\% centrality region 
indicated, compared to the calculation of the distribution given by 
Eq.~21 for $N_{\rm part}$=254 and 
$N_{\rm coll}=672$ 
corresponding to 10\%--15\% centrality.}
\label{fig:crazy}
\end{figure}

\begin{figure}[htb] 
\includegraphics[width=1.0\linewidth]{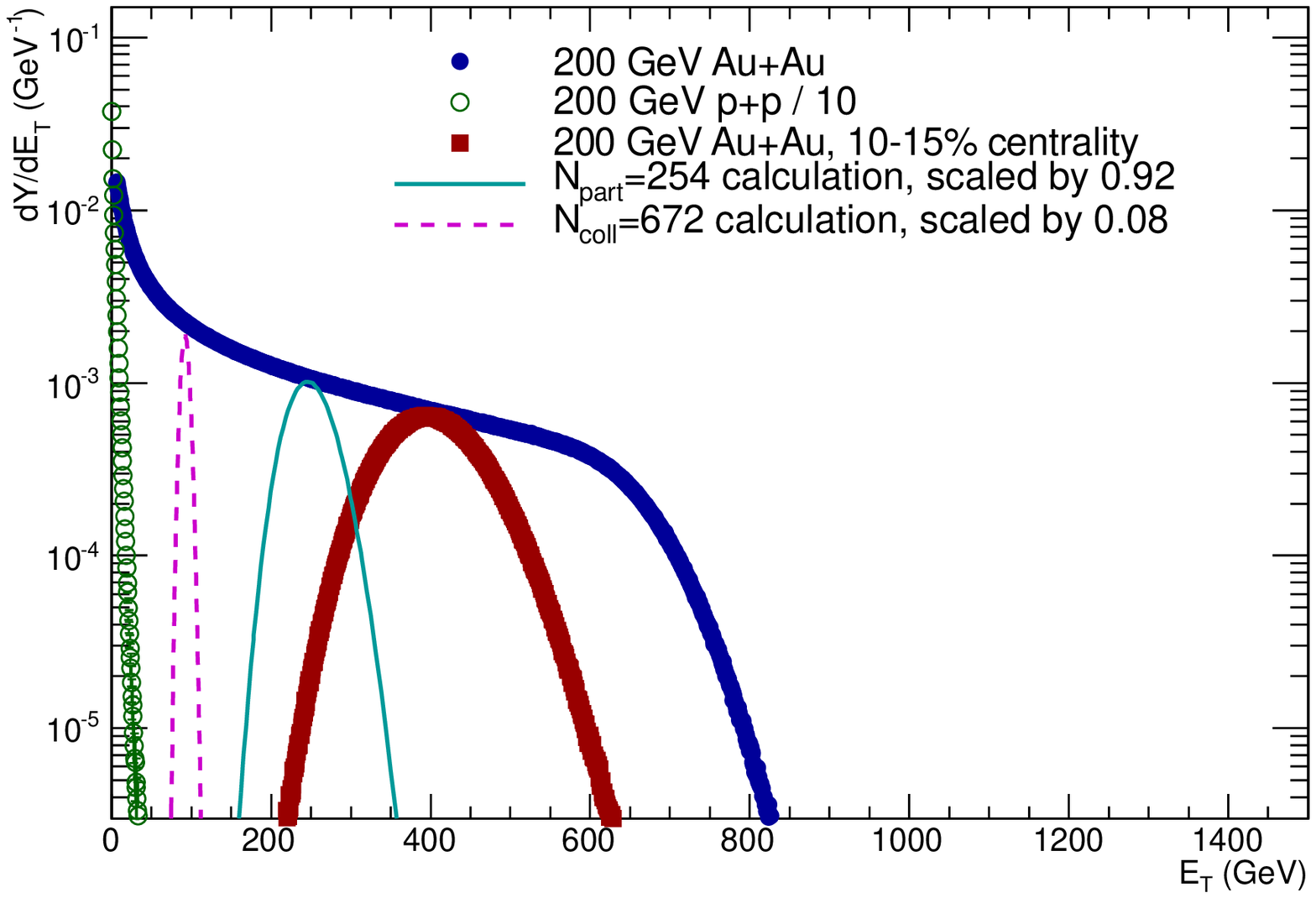} 
 \caption{(Color online) 
Au$+$Au measurement of $d\Et/d\eta$, with 10\%--15\% centrality region 
indicated, compared to the calculation of the distribution given by 
Eq.~21 for $N_{\rm part}$=254 and $N_{\rm coll}=672$ 
corresponding to 10\%--15\% centrality, where the distributions have been 
scaled in \Et by 0.92 and 0.08, respectively.}
      \label{fig:crazy2}
\end{figure}

\subsection{Additional Systematic Uncertainties} 

The probability $p_0$ of detecting zero \Et in the central detector for an 
N+N or other elementary collision plays a major role in this analysis. The 
predominant systematic uncertainty comes from the BBC cross section 
measurement (Eq.~\ref{eq:1-p0}) which leads to a total systematic 
uncertainty on $1-p_0$ of 10.1\% for a $p$$+$$p$ collision as indicated in 
Table~\ref{table:Gammafitpp}. The systematic uncertainty is propagated by 
varying $1-p_0$ from 0.647 to 0.712 and 0.582, $\pm 1$ standard deviation, 
from Eq.~\ref{eq:1-p0} for standard $\Etemc<2$ GeV ($\Et<13.3$ GeV) 
$p$$+$$p$ data and repeating all the calculations (to be shown in 
Sec.~\ref{sec:Final}). Also all the fits were redone with the $\Etemc<4$ 
GeV ($\Et<26.6$ GeV)  $p$$+$$p$ data, and the calculations were again all 
repeated, with a small effect (Fig.~\ref{fig:syscheck}).

Another important issue must be mentioned in the comparison of the 
calculations to the measurements. The calculations are per $A$+$B$ 
collision (corrected for BBC efficiency) while the data are per BBC count 
and are not corrected for the BBC efficiency. This correction is 
complicated for both $d$$+$Au and Au$+$Au, but larger for Au$+$Au due to the 
more severe BBC requirement.  To get an idea of the size of the effect, 
Fig.~\ref{fig:v12p5b} shows the Au$+$Au data and the NQP calculation of 
Fig.~\ref{fig:all} on the same \Et scale as in $d$$+$Au 
collisions (see Fig.~\ref{fig:f4}). 
The inefficiency in the data compared to the Au$+$Au calculation is 
negligible for $\Et\geq7$ GeV, as shown, which should be less severe for 
$d$$+$Au and therefore does not affect the conclusion from 
Fig.~\ref{fig:dAufits} that rejects the AQM model in favor of the NQP 
model.

\section{Final Results and discussion}\label{sec:Final}

The principal results were given in Figs.~\ref{fig:all} and 
\ref{fig:dAufits}.  The final results are now presented in 
Fig.~\ref{fig:dAuAufinal} including the systematic uncertainties. In 
Fig.~\ref{fig:dAuAufinal}a, the Au$+$Au $\Et\equiv d\Et/d\eta|_{y=0}$ 
distribution is shown compared to the NQP calculations using the central 
$1-p_0=0.647$ and $\pm 1\sigma$ variations of $1-p_0=0.582,0.712$ for the 
probability of getting zero \Et on a $p$$+$$p$ collision, which correspond 
to $\epsilon_{\rm NQP}=0.659, 0.603,0.716$ respectively. Both the shape 
and magnitude of the calculation with the NQP model are in excellent 
agreement with the Au$+$Au measurement. The upper edge of the calculation 
using the central $1-p_0$ is essentially on top of the measured \Et 
distribution, well within the principal $\pm 10\%$ systematic uncertainty 
shown, while the AQM model (recall Fig.~\ref{fig:all}) was another 12\% 
lower due to the nonzero $p_0$ in $p$$+$$p$ collisions in this measurement 
which leads to different efficiencies of a quark participant and a color 
string.

In Fig.~\ref{fig:dAuAufinal}b the $d$$+$Au \Et distribution is shown with 
the central $1-p_{0_{\rm NQP}}$ and the $\pm 1\sigma$ variations.  The NQP 
calculation closely follows the $d$$+$Au measurement in shape and in 
magnitude over a range of a factor of 1000 in cross section, while as 
previously seen in Fig.~\ref{fig:dAufits}, the AQM calculation disagrees 
both in shape and magnitude, with nearly a factor of 2 less \Et emission. 
A new independent check of the NQP model is the observation that the 
$\mean{d\Et/d\eta}/N_{qp}=0.617\pm0.023$ GeV calculated from the linear 
fit (Fig.~\ref{fig:detdetaNquark}) of the Au$+$Au measurement as a 
function of centrality is equal (within $<1$ standard deviation) to the 
value $\mean{\Et}^{\rm true}_{qp}=0.655\pm 0.066$ GeV derived for a 
quark-participant from the deconvolution of the $p$$+$$p$ \Et distribution 
(Table~\ref{table:Gammafits}).

The availability of the $p$$+$$p$ baseline \Et distribution together with 
the Au$+$Au distribution allows a test of how the representation of 
${d\Nch/d\eta}$ or ${d\Et/d\eta}$ as a function of centrality by 
this rewrite of Eq.~\ref{eq:crazy}~\cite{Adcox:2000sp,WangGyulassyPRL86,KharzeevNardiPLB507}:
\begin{eqnarray}
 {d\Et^{AA}/d\eta}
&=&[(1-x)\,\mean{\Npart} (d\Et^{pp}/d\eta)/2 \\ \nonumber
&+&x\,\mean{\Ncoll} (d\Et^{pp}/d\eta)],
\label{eq:repeat}
\end{eqnarray}
which works for the average values, could be applied to the distributions.


Figure~\ref{fig:v12p4bot} compares the Au$+$Au data to the \Ncoll and 
\Npart-WNM calculations, including the efficiencies. One thing that is 
immediately evident from Fig.~\ref{fig:v12p4bot} is that if 
Eq.~\ref{eq:crazy},21 were taken to represent the weighted 
sum of $(1-x)\ \times$ the WNM-\Npart curve + $x\ \times$ the \Ncoll curve 
with $x\approx 0.08$~\cite{Adcox:2000sp,PHOBOSPRC70}, then the 
representation of ${d\Et/d\eta}$ by Eq.~\ref{eq:crazy},21, 
which may seem reasonable for the average values, makes no sense for the 
distribution.
    
To further emphasize this point, shown in Fig.~\ref{fig:crazy} is the 
calculation of the distribution given by 
Eq.~\ref{eq:crazy},21 for 10\%--15\% centrality, namely the 
sum of the \Npart distribution for $\mean{N_{\rm part}}=254$, weighted by 
(1-$x$), and the \Ncoll distribution for $\mean{N_{\rm coll}}=672$ 
weighted by $x$, compared to the
measured Au$+$Au distribution for 10\%--15\% percentile centrality 
region.~\footnote{The curves in Fig.~\protect\ref{fig:crazy} are actually 
for $254\times(\epsilon_{N_{\rm part}}=0.804)=204$ convolutions of 
$f_1^{N_{\rm part}}$ and $672\times(\epsilon_{N_{\rm coll}}=0.647)=435$ 
convolutions of the $p$$+$$p$ measured reference distribution, 
$f_1^{N_{\rm coll}}(\Et)$ following Eq.~\protect\ref{eq:true-refalln}.} 
Although it is reasonable that the weighted sum of the averages of the 
\Ncoll and \Npart distributions could equal the average of the measured 
${d\Et/d\eta}$ distribution for 10\%--15\% centrality, the weighted sum of 
the actual \Ncoll and \Npart distributions would look totally unreasonable 
and nothing like the measured ${d\Et/d\eta}$ distribution cut on 
centrality. Thus Eq.~\ref{eq:crazy} can not be interpreted as the weighted 
sum of the \Ncoll and \Npart distributions. Furthermore, as shown in 
Fig.~\ref{fig:crazy2}, neither can Eq.~\ref{eq:crazy} be interpreted as 
the sum of the \Ncoll and \Npart distributions scaled in \Et by the 
factors $x$ and $1-x$ respectively. Hence it does not seem that 
Eq.~\ref{eq:crazy} can be computed in an extreme independent model.
	 
Recent experiments at the Large Hadron Collider, the 
ATLAS experiment in particular~\cite{ATLASPLB707}, have shown that 
computing Eq.~\ref{eq:crazy} on an event-by-event basis as a nuclear 
geometry distribution in a standard Glauber calculation, agrees very well 
with their measured \Et distribution in the pseudorapidity range 
$3.2<|\eta|<4.9$ at \sqsn=2.76 TeV Pb$+$Pb collisions. Similar results 
were obtained by ALICE~\cite{ALICEPRC88}. This confirms the observation 
noted previously (Sec.~\ref{sec:results}) that the success of the two 
component model is not because there are some contributions proportional 
to \Npart and some proportional to \Ncoll, but rather because a particular 
linear combination of \Npart and \Ncoll turns out to be an empirical proxy 
for the nuclear geometry of the number of constituent-quark participants, 
\Nqp in A+A collisions.

\section{Summary}
\label{sec:summary}

To summarize, the midrapidity transverse energy distributions, 
$d\Et/d\eta$, have been measured for \sqsn=200 GeV $p$$+$$p$ and $d$$+$Au 
collisions, and for Au$+$Au collisions at \sqsn=200, 130, and 62.4 GeV.  
As a function of centrality, the $\mean{d\Et/d\eta}$ measured in Au$+$Au 
collisions at all three collision energies exhibit a nonlinear increase 
with increasing centrality when expressed as the number of nucleon 
participants, \Npart. When expressed in terms of the number of 
constituent-quark participants, $N_{qp}$, the $\mean{d\Et/d\eta}$ 
increases linearly with $N_{qp}$. Several Extreme Independent models of 
particle production have been compared to the data, including calculations 
based upon color-strings (the Additive Quark Model, AQM) and the 
constituent-Quark Participant model (NQP). When compared to data from 
symmetric systems (Au$+$Au and $p$$+$$p$), these two models cannot generally 
be distinguished from each other. In the present measurement, the 
different detection efficiency for a quark-participant and color string in 
the two cases allows a separation, with the NQP model favored. However, 
when compared to data from the asymmetric $d$$+$Au system, the $d$$+$Au 
measurement clearly rejects the AQM model and agrees very well with the 
NQP model. This implies that transverse energy production at midrapidity 
in relativistic heavy ion collisions is well described by particle 
production based upon the number of constituent-quark participants. 
Additional support for this conclusion is that the ansatz, 
$[(1-x)\,\mean{\Npart}/2 +x\,\mean{\Ncoll}]$, which has been used 
successfully to represent the nonlinearity of $\mean{d\Et/d\eta}$ as a 
function of \Npart, turns out to be simply a proxy for $\mean{\Nqp}$ in 
A+A collisions and does not represent a hard-scattering component in \Et 
distributions.

\section*{ACKNOWLEDGMENTS}

We thank the staff of the Collider-Accelerator and Physics
Departments at Brookhaven National Laboratory and the staff of
the other PHENIX participating institutions for their vital
contributions.  We acknowledge support from the 
Office of Nuclear Physics in the
Office of Science of the Department of Energy, the
National Science Foundation, Abilene Christian University
Research Council, Research Foundation of SUNY, and Dean of the
College of Arts and Sciences, Vanderbilt University (U.S.A),
Ministry of Education, Culture, Sports, Science, and Technology
and the Japan Society for the Promotion of Science (Japan),
Conselho Nacional de Desenvolvimento Cient\'{\i}fico e
Tecnol{\'o}gico and Funda\c c{\~a}o de Amparo {\`a} Pesquisa do
Estado de S{\~a}o Paulo (Brazil),
Natural Science Foundation of China (P.~R.~China),
Centre National de la Recherche Scientifique, Commissariat
{\`a} l'{\'E}nergie Atomique, and Institut National de Physique
Nucl{\'e}aire et de Physique des Particules (France),
Bundesministerium f\"ur Bildung und Forschung, Deutscher
Akademischer Austausch Dienst, and Alexander von Humboldt Stiftung (Germany),
Hungarian National Science Fund, OTKA (Hungary), 
Department of Atomic Energy (India), 
Israel Science Foundation (Israel), 
National Research Foundation and WCU program of the 
Ministry Education Science and Technology (Korea),
Physics Department, Lahore University of Management Sciences (Pakistan),
Ministry of Education and Science, Russian Academy of Sciences,
Federal Agency of Atomic Energy (Russia),
VR and Wallenberg Foundation (Sweden), 
the U.S. Civilian Research and Development Foundation for the
Independent States of the Former Soviet Union, the US-Hungarian
NSF-OTKA-MTA, and the US-Israel Binational Science Foundation.




\end{document}